\documentclass[useAMS,usenatbib]{mn2e} 
\usepackage{graphics, graphicx} 

\title[Absolute polarisation position angle profiles of southern
  pulsars at 1.4 and 3.1 GHz]{Absolute polarisation position angle
  profiles of southern pulsars at 1.4 and 3.1 GHz}

\author[Karastergiou et al.]  {A.~Karastergiou$^1$ \& S.~Johnston$^{1,2}$\\
$^1$School of Physics, University of Sydney, NSW 2006, Australia\\
$^2$Australia Telescope National Facility, CSIRO, P.O. Box 76, Epping,
NSW 1710, Australia.\\ }

\date{Released 2004 Xxxxx XX}
 
\pagerange{\pageref{firstpage}--\pageref{lastpage}} \pubyear{2004} 
 
\def\LaTeX{L\kern-.36em\raise.3ex\hbox{a}\kern-.15em 
    T\kern-.1667em\lower.7ex\hbox{E}\kern-.125emX} 
 
\begin{document} 
 
\label{firstpage} 
 
\maketitle 
 
\begin{abstract} 
We present here a direct comparison of the polarisation position angle
(PA) profiles of 17 pulsars, observed at 1.4 and 3.1 GHz. Absolute PAs
are obtained at each frequency, permitting a measurement of the
difference in the profiles. By doing this, we obtain more precise
rotation measure (RM) values for some of the pulsars in the current
catalogue. We find that, apart from RM corrections, there are small,
pulse longitude dependent differences in PA with frequency. Such
differences go beyond the interpretation of a geometrical origin. We
describe in detail the PA evolution between the two frequencies and
discuss possible causes, such as orthogonal and non-orthogonal
polarisation modes of emission. We also use the PA and total power
profiles to estimate the difference in emission height at which the
two frequencies originate. In our data sample, there are changes in
the relative strengths of different pulse components, especially
overlapping linearly polarised components, which coincide with
intrinsic changes of the PA profile, resulting in interesting PA
differences between the two frequencies.

\end{abstract} 
 
\section{Introduction}
Shortly after the original discovery of pulsars, \citet{rc69a}
developed a model of emission which has subsequently been used to
interpret pulsar polarisation. According to this model, pulsar
radiation is polarised along, or orthogonal to, the open magnetic
field lines of the strong dipolar magnetic field that surrounds the
star. As the beam of the pulsar sweeps past the line of sight of the
observer, the position angle (PA) of the linear polarisation rotates,
resulting in the well known S-shaped curves of PA versus pulse phase
seen in many pulsars. The angles which determine the sweep of the PA
are the inclination angle $\alpha$ between the magnetic and rotational
axes and the impact parameter $\beta$, which is the smallest angle
between the locus of the line of sight on the pulsar beam and the
magnetic axis. This rotating vector model (RVM) has been used in some
cases to infer the geometrical angles of pulsars from polarisation
observations.

Despite the initial success of the RVM, numerous attempts to fit
polarimetric data demonstrated that in many cases, the parameters of
the fit were poorly constrained, mainly due to 
the restricted longitude over which pulsars emit. For instance, in the case of
\citet{bcw91}, who are primarily concerned with the point of steepest
gradient of the PA swing, the fitted parameters $\alpha$ and $\beta$
at 430 and 1418 MHz rarely agree. The authors attribute this to very
large error estimates, often as large as hundreds of degrees. The same
is seen in another set of published RVM fits at multiple radio
frequencies \citep{hx97a}. \citet{ml04} recently presented RVM fits
for 6 pulsars, to data obtained at multiple frequencies (4 to 6
frequencies for each pulsar). Similarly to \citet{bcw91}, the fitted
parameters (especially $\alpha$) generally do not agree between the
different frequencies, despite the claim of the authors to the
contrary. In a remarkably complete treatment, \citet{ew01} summarise
the problem and demonstrate the difficulties in obtaining a unique RVM
solution.  They attempt RVM fits on their data, obtaining reliable
results for only 10 out of the 70 pulsars in their sample. Even in the
limited cases, the fitted parameters are not always in agreement with
previously published results.

Early observations also showed that average pulse profiles become
narrower with increasing frequency. The model of \citet{rs75} suggests
that the frequency of emission is related to the local plasma density,
which drops off at higher altitudes from the surface of the star. This
leads to a natural radius-to-frequency mapping (RFM), with higher
frequencies originating closer to the pulsar surface. In the dipolar
magnetic field structure, the open field lines spread out at greater
altitudes, explaining the broader widths of profiles at lower
frequencies \citep{cor78}. Although the exact height at which a
particular frequency is emitted depends on the nature of the emission
mechanism \citep{mel00a}, some sort of RFM appears to underly the
observations.

To first order, the PA at a given pulse longitude is fixed,
irrespective of the emission height. The consequences of RFM on the PA
lie mainly in second order effects, such as relativistic effects
within the pulsar light cylinder. \citet{bcw91} showed that due to
such effects, the centroid of the PA swing is delayed with respect to
the centroid of the total intensity by an amount proportional to the
radius at which the radiation originates, providing a method to
estimate relative emission heights in the magnetosphere. Such
calculations agreed well with the heights inferred by the changing
widths of average profiles.

In a comparison of data from PSR B0329+54 at two
frequencies,\citet{ms98a} showed that the position of the pair of
outer components changes relative to the central component, due to
effects of retardation and aberration.  A method for estimating
emission altitudes was subsequently proposed by \citet{gg01}, who
exploit the fact that components originating close to the magnetic
axis are emitted from much lower altitudes than pairs of components
that flank them. In pulsar profiles with an appropriate configuration,
the relativistic phase shift between the central (core) component and
the middle of the outer component pair provides the only information
necessary to compute emission heights, as demonstrated in a refinement
of the method of \citet{gg01} by \citet{drh04}. The emission altitudes
using this relativistic method are generally larger than those of the
\citet{bcw91} method. In practice, the main difficulty of estimating
emission heights in this way relates to the limited pulsar profiles
with a central component and outer components that can be
unambiguously identified as a pair.

There are several reasons that cause observed PAs to diverge from the
RVM. The first and probably best documented phenomenon is orthogonally
polarised modes (OPM) in pulsar emission
\citep[e.g.][]{mth75,brc75,crb78}. These result in parts of the PA
profile being $90\degr$ offset to the main swing, an effect which can
be identified and accounted for. Also, OPMs are frequency dependent
\citep[e.g.][]{scr+84,kkj+02}, which results in different PA profiles
at different frequencies, the higher frequencies showing more
occurrences of $90\degr$ discontinuities.

An additional source of the PA dependence on frequency may originate
from non-orthogonal polarisation modes, identified first by
\citet{br80}. \citet{rbr+04} recently showed that the different
spectral behaviour of quasi-orthogonal modes in PSR B2016+28 results
in a change in the PA swing over a very narrow frequency range.
\citet{kjm05} claim that the spectral behaviour of the total power and
linear polarisation in different components can also be attributed to
different spectra of the orthogonal modes. The possibility of variable
PA profiles at different frequencies then naturally arises.

\citet{mic91} considered Faraday rotation in the magnetospheres of
pulsars as the cause of complexity in the PA profiles of
pulsars. However, he argued that the physical grounds for this process
are lacking and subsequently no attempt was made to observationally
identify this phenomenon.

All of the above mechanisms that distort the simple geometric nature
of the RVM, also have some form of inherent dependence on frequency.
We therefore proceeded to investigate this by presenting for the first
time direct comparisons of the PA swings of 17 pulsars between 1.375
and 3.1 GHz. In doing so, we document the nature of the frequency
evolution of the PA, which we discuss in the context of the above
physical processes and models.
The purpose of our analysis is to specifically avoid RVM fits to our
data and geometrical interpretations which rely upon these fits. On
the contrary, instead of comparing the PA profiles to a badly
constrained curve, we compare them as precisely as possible to each
other in an unbiased approach to study PA changes. We believe this
approach is warranted given the nature of most PA profiles in our data
and the difficulties in fitting RVM curves over very small longitude
ranges.

In Section 2 we describe the observations. In Section 3 we present a
method by which we align the profiles at the two different frequencies
using the absolute PA values, which also permit us to determine very
precise RMs. In Section 4 we discuss the polarisation profiles and
compare them to previous profiles at other frequencies. In the final
parts of this paper, we discuss the differences between the profiles
at the two frequencies and possible interpretations in the context of
the above models for the origin and frequency evolution of PA
profiles.

\section{The observations}

The observations were carried out using the Parkes radio telescope on
2004 August 31 and September 1. We used the H-OH receiver at a central
frequency of 1.375 GHz with a bandwidth of 256~MHz and the 10/50~cm
receiver at a central frequency of 3.1~GHz with a bandwidth of
512~MHz.  In both cases, the backend correlator subdivided the total
bandwidth into 1024 frequency channels and also recorded 1024
longitude bins per pulse period.  The receivers have orthogonal linear
feeds and also have a pulsed cal signal which is injected at a
position angle of 45\degr\ to the two feed probes.

The pulsars were observed for 30 minutes at each frequency. Prior to
the pulsar observation a 3~min observation of the pulsed cal was made.
The data were written to disk in FITS format for subsequent off-line
analysis. Data analysis was carried out using the PSRCHIVE software
package \citep{hvm04}.
\begin{table*}
\begin{tabular}{cccccccl}
\multicolumn{2}{c}{Name} & Period & \multicolumn{2}{c}{W$10$} &
\multicolumn{2}{c}{RM} & RM reference \\
J2000     & B1950  & ms     & $\nu_1$     & $\nu_2$         &
\multicolumn{2}{c}{rad/m$^2$}& \\
          &        &        & deg         & deg         & New       & Previous   &     \\   
\hline
J0738$-$4042& B0736-40 & 375    & 32          & 30          & 14$\pm0.5$&   14  & \citet{tml93}\\
J0742$-$2822& B0740-28 & 167    & 16          & 16          &150$\pm0.5$&  156  & \citet{hmq99} \\
J0835$-$4510& B0833-45 & 89     & 14          & 16          & 32$\pm0.5$&   38  & \citet{hmm+77} \\
J0837$-$4135& B0835-41 & 752    & 5           & 6           &146$\pm1.0$&  136  & \citet{tml93} \\
J0942$-$5552& B0940-55 & 664    & 26          & 29          &$-$62$\pm0.8$&  $-$62  & \citet{tml93} \\
J1056$-$6258& B1054-62 & 422    & 32          & 32          &$-$1 or 6$\pm0.5$&    4  & \citet{cmh91} \\
J1157$-$6224& B1154-62 & 401    & 45          & 48          &507$\pm2.5$&  508  & \citet{tml93} \\
J1243$-$6423& B1240-64 & 388    & 7           & 8           &162$\pm1.0$&  158  & \citet{tml93} \\
J1326$-$5859& B1323-58 & 478    & 17          & 16          &$-$586$\pm0.7$& $-$580  & \citet{cmh91} \\
J1327$-$6222& B1323-62 & 530    & 11          & 11          &$-$319$\pm0.8$&       &  \\
J1356$-$6230& B1353-62 & 456    & 32          & 27          &$-$580$\pm1.7$&       &  \\
J1359$-$6038& B1356-60 & 127    & 14          & 14          &39 or 37$\pm0.5$&   33  & \citet{hmq99} \\
J1602$-$5100& B1558-50 & 864    & 11          & 13          &79$\pm1.3$ &   72  & \citet{tml93} \\
J1644$-$4559& B1641-45 & 455    & 14          & 11          &$-$618$\pm0.5$& $-$611  & \citet{tml93} \\
J1752$-$2806& B1749-28 & 562    & 8           & 8           &91$\pm3.0$ &   96  & \citet{hl87} \\
J1807$-$0847& B1804-08 & 164    & 28          & 28          &171$\pm1.8$&  166  & \citet{hl87} \\
J1825$-$0935& B1822-09 & 769    & 22          & 21          &68$\pm0.7$ &   65  & \citet{hl87} \\
\end{tabular}
\caption{The list of pulsars and their parameters. $\nu_1$ and $\nu_2$
  correspond to 1.375 and 3.1 GHz respectively.}
\end{table*}

We used standard techniques to calibrate the relative gains and phases
of the two linear feed probes. The pulsed cal is injected at 45\degr\
to both probes so that its intrinsic signal can be assumed to be 100
per cent polarised in Stokes $U$.  Gain corrections can be made by
comparing the observed Stokes $Q$ with Stokes $I$ and phase
corrections can be made by comparing the observed Stokes $U$ with
Stokes $V$ on a channel by channel basis in the output data.
Following this calibration $Q$ and $U$ can be further rotated to
correct for the angle which the feed probes make to true north and for
the parallactic angle of the source at the time of the observation. At
the end of this procedure we have the intrinsic PA of the radiation
just above the feed at a particular frequency.  PA values are only
computed when the linear polarisation exceeds the noise by a factor of
5.

To obtain the PA at the pulsar, the Faraday rotation both through the
Earth's ionosphere and the interstellar medium must be corrected
for. This is done in a two stage process. First, we fit for the RM
across the 256~MHz band at 1369~MHz.  This yields an RM with an error
of order $\pm$5~rad~m$^{-2}$.  The PAs at both 1.375 and 3.1 GHz can
be rotated by an amount given by
\begin{equation}
\theta = \frac{c^2 \,\,\, {\rm RM}}{\nu^2}
\end{equation}
where $c$ is the speed of light and $\nu$ the observing frequency. A
more precise RM can then be produced by comparing the PAs at both
frequencies, using a least squares fitting technique as described in
more detail below.

\section{Data analysis}

A new approach is used in this paper to align profiles at 1.375 and 3.1
GHz. Initially, we slide one PA profile over the other in the
horizontal, pulse-longitude direction, in steps of one bin of data. At
every step of shift $dx$, we measure the average offset $d$PA between
the two PA profiles
\begin{equation}
d{\rm PA}_{dx}=\frac{1}{N}\sum_{i=1}^{N} {\rm PA}1_i-{\rm PA}2_i ,
\end{equation}  
where $N$ is the overlapping number of data bins which have a PA
measurement at both frequencies, $i$ the index of each such bin and
PA1 and PA2 the value of the PA at that bin at 1.375 and 3.1 GHz
respectively. Then, we calculate a normalised chi-square between the
two sets of data, considering their average PA offset, according to
the equation
\begin{equation}
\chi^2_{dx}=\frac{1}{N}\sum_{i=1}^{N}\frac{({\rm PA}1_i-{\rm PA}2_i-d{\rm
PA}_{dx})^2}{2(\sigma_{{\rm PA}1,i}^2+\sigma_{{\rm PA}2,i}^2)} ,
\end{equation}
where $\sigma_{{\rm PA}1}$ and $\sigma_{{\rm PA}2}$ are the
uncertainties in the PA measurements. As shown later, the instances of
orthogonal PAs at the two frequencies are few and do not influence the
value of $\chi^2_{dx}$. Despite this, if any instantaneous difference
in the PA is greater than $45\degr$, we assign the PA difference the
complementary value to $90\degr$, thus avoiding great dPA values in
the case of orthogonal PAs at the two frequencies. We then shift the
profile by one bin and repeat the process, thus yielding a table of
$\chi^2$ as a function of the shift $dx$. The value of $dx$ that
corresponds to $\chi^2_{min}$ is chosen as the shift in
pulse-longitude to align the two profiles, and the corresponding
average $d$PA is used to apply a vertical shift.

The vertical shift that is required to align the profiles can be
achieved with a correction in the RM value and therefore by
calculating the mean $d$PA offset we are computing a very accurate
RM. The error on the RM then corresponds to the error in the mean
$d$PA, which is
\begin{equation}
\sigma_{d\rm PA}=\frac{1}{N}\sum_{i=1}^{N}\sqrt{\sigma_{{\rm
PA}1,i}^2+\sigma_{{\rm PA}2,i}^2}
\end{equation}

The contributions to the measured values of RM arising in the
ionosphere have been estimated by integrating a time-dependent model
of the ionospheric electron density and geomagnetic field through the
ionosphere along the sight line between the telescope and the
pulsar. The average contribution is found to be $\sim -1$~rad~m$^{-2}$.

\section{Results}

\begin{figure*}
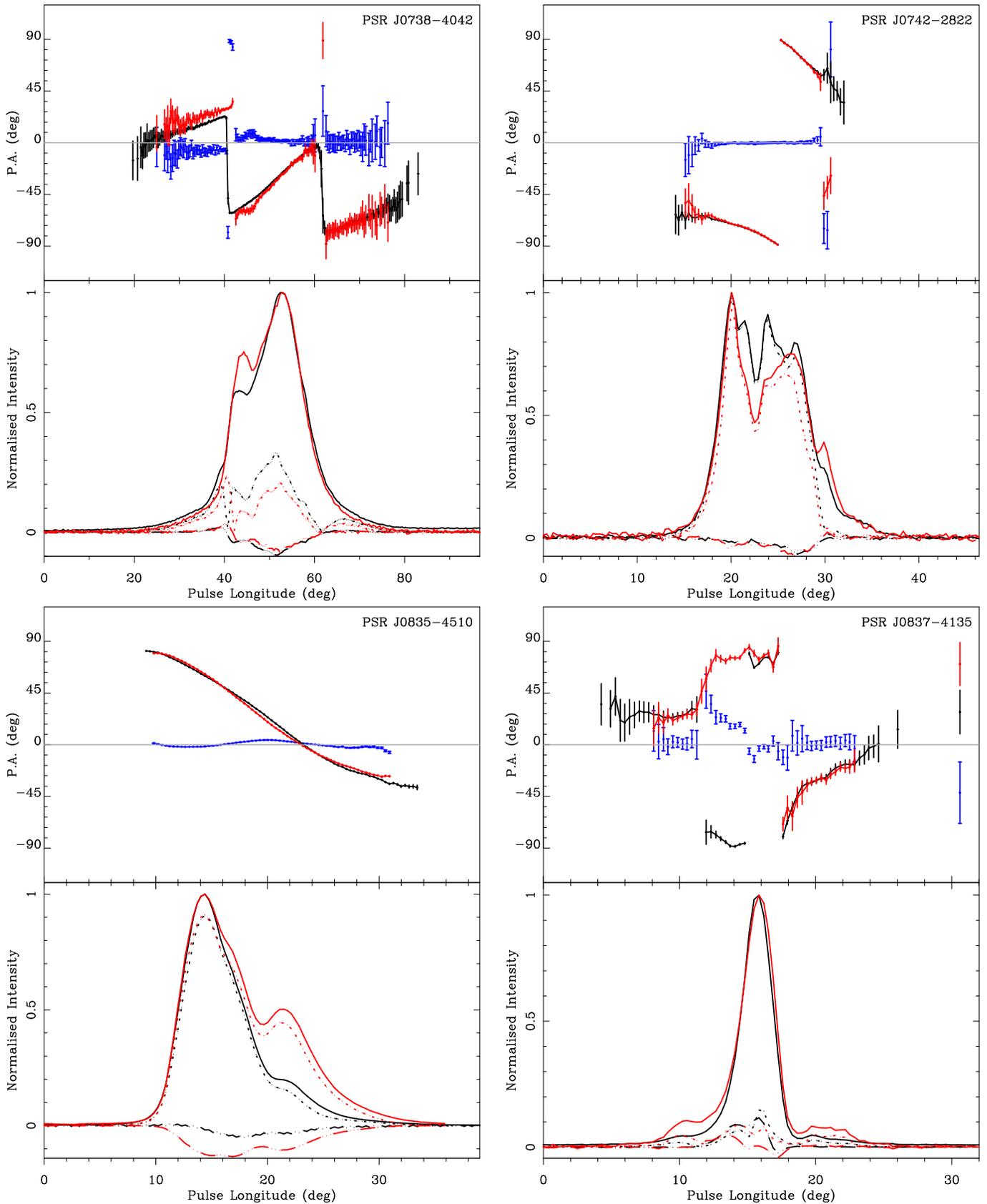

\begin{center}
\begin{tabular}{cc}
\resizebox{0.5\hsize}{!}{\includegraphics[angle=0]{J0738-4042.ps}}&
\resizebox{0.5\hsize}{!}{\includegraphics[angle=0]{J0742-2822.ps}}\\
\resizebox{0.5\hsize}{!}{\includegraphics[angle=0]{J0835-4510.ps}}&
\resizebox{0.5\hsize}{!}{\includegraphics[angle=0]{J0837-4135.ps}}\\
\end{tabular}
\end{center}
\caption{Overlaid polarisation profiles at 1.375 (black lines) and 3.1
  GHz (red). In the lower panel of each figure, the solid lines denote
  total power, doted lines linear polarisation and dash-dotted lines
  circular polarisation. In the upper panel, the absolute PAs are
  shown for 1.375 and 3.1 GHz, following the same colour
  convention. Also, the error-bars with riser lines (blue) denote the
  difference between the two PAs, where there is a measurement at both
  frequencies.}
\end{figure*}
\addtocounter{figure}{-1}
\begin{figure*}
\begin{center}
\begin{tabular}{cc}
\resizebox{0.5\hsize}{!}{\includegraphics[angle=0]{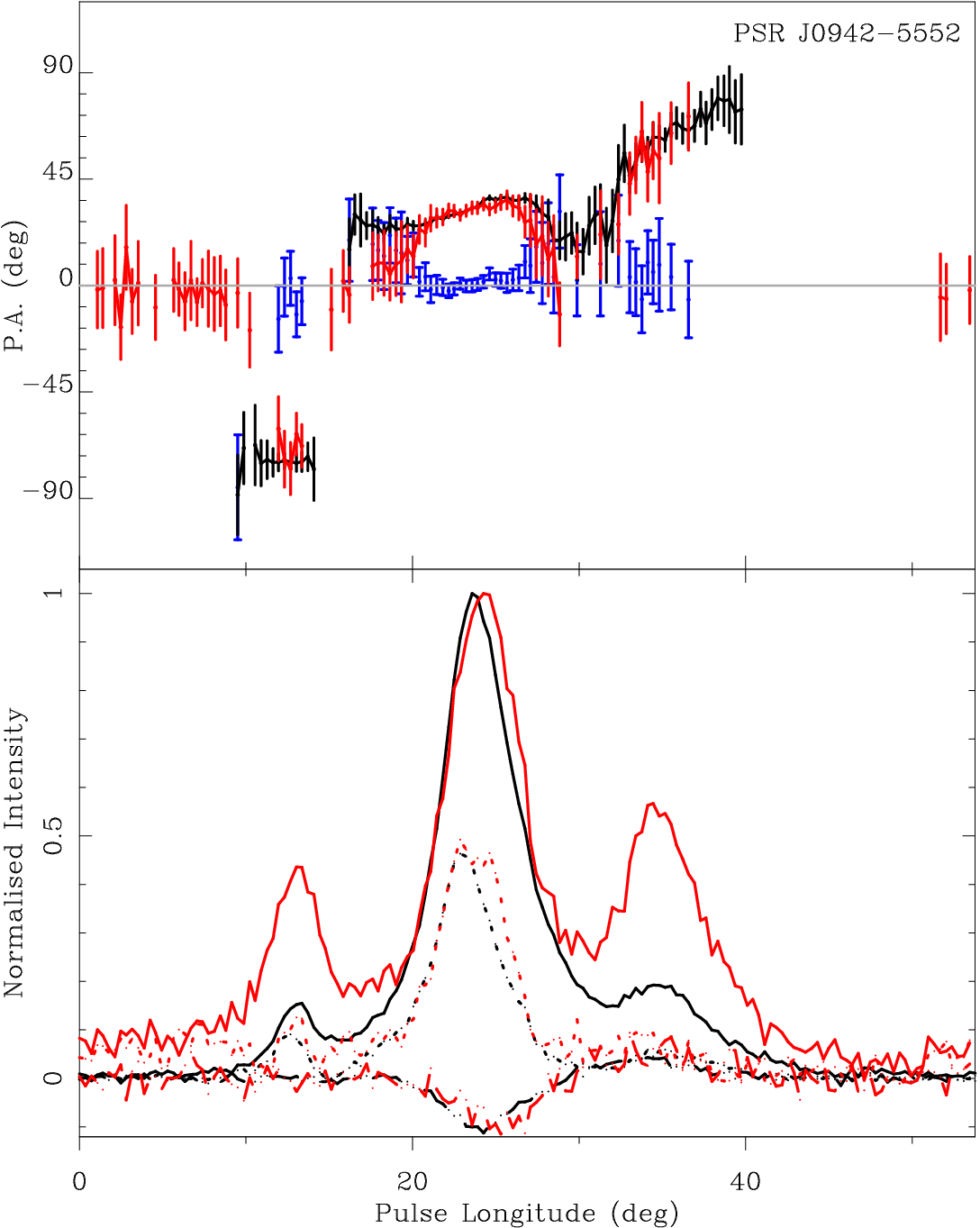}}& 
\resizebox{0.5\hsize}{!}{\includegraphics[angle=0]{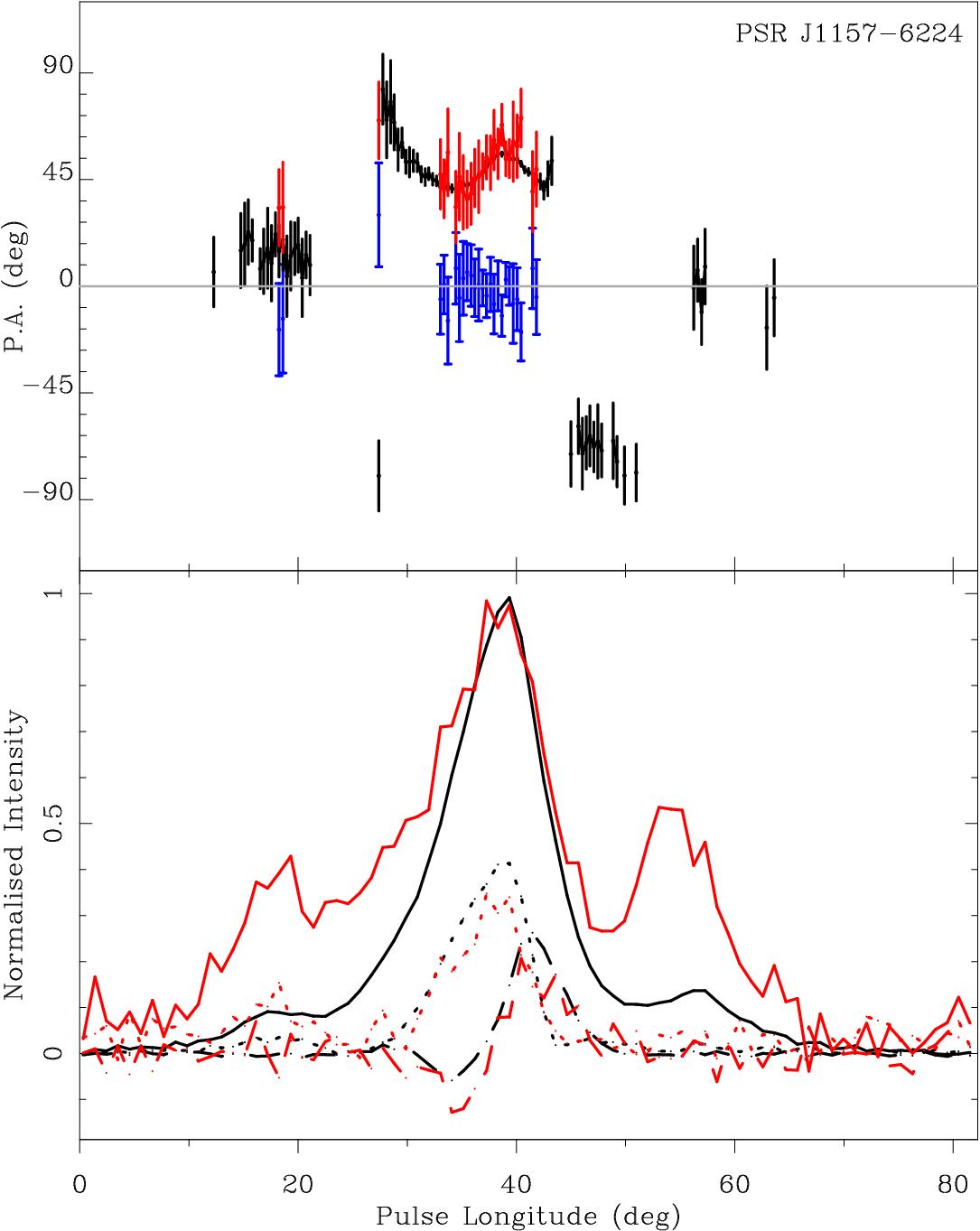}}\\ 
\resizebox{0.5\hsize}{!}{\includegraphics[angle=0]{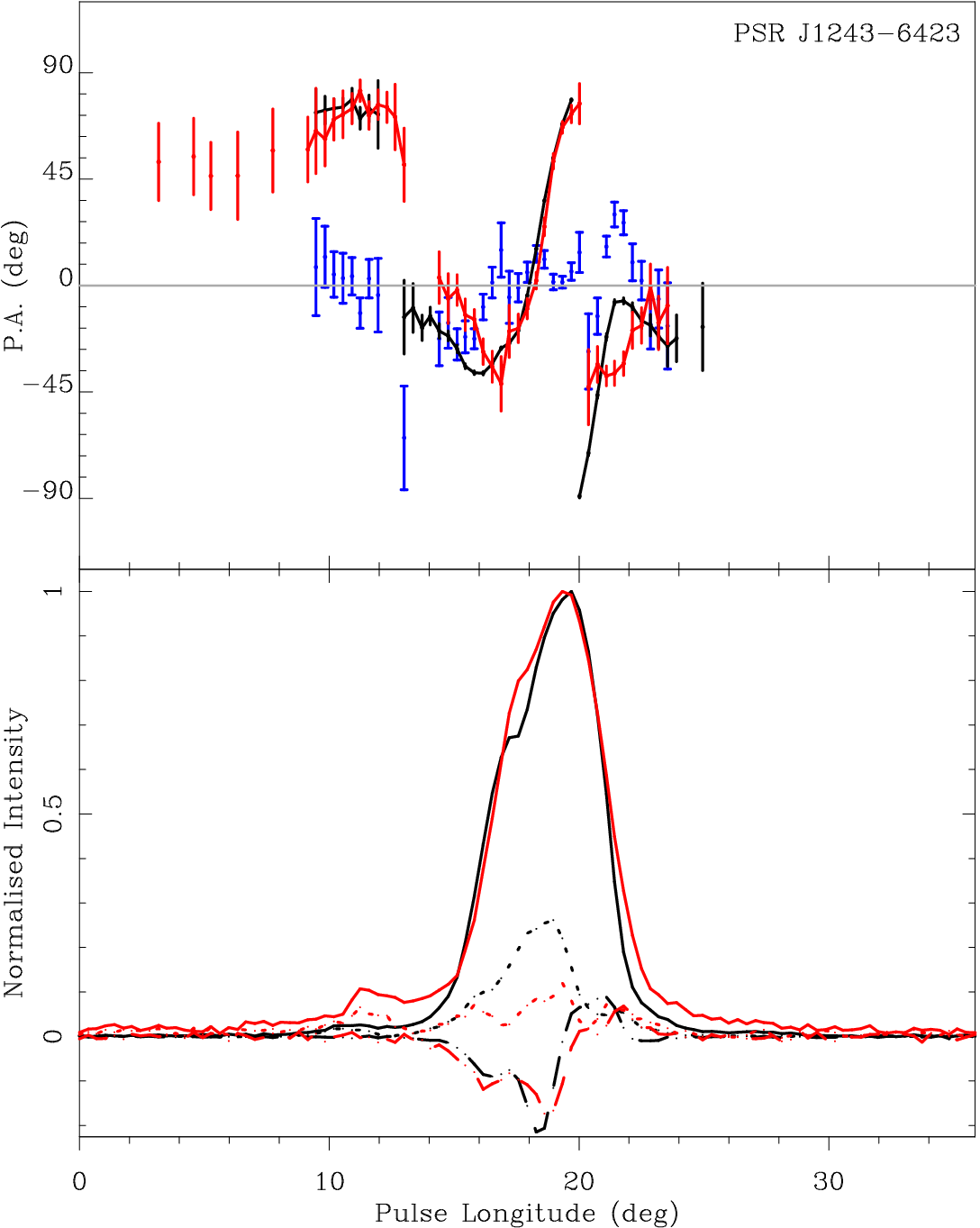}}&
\resizebox{0.5\hsize}{!}{\includegraphics[angle=0]{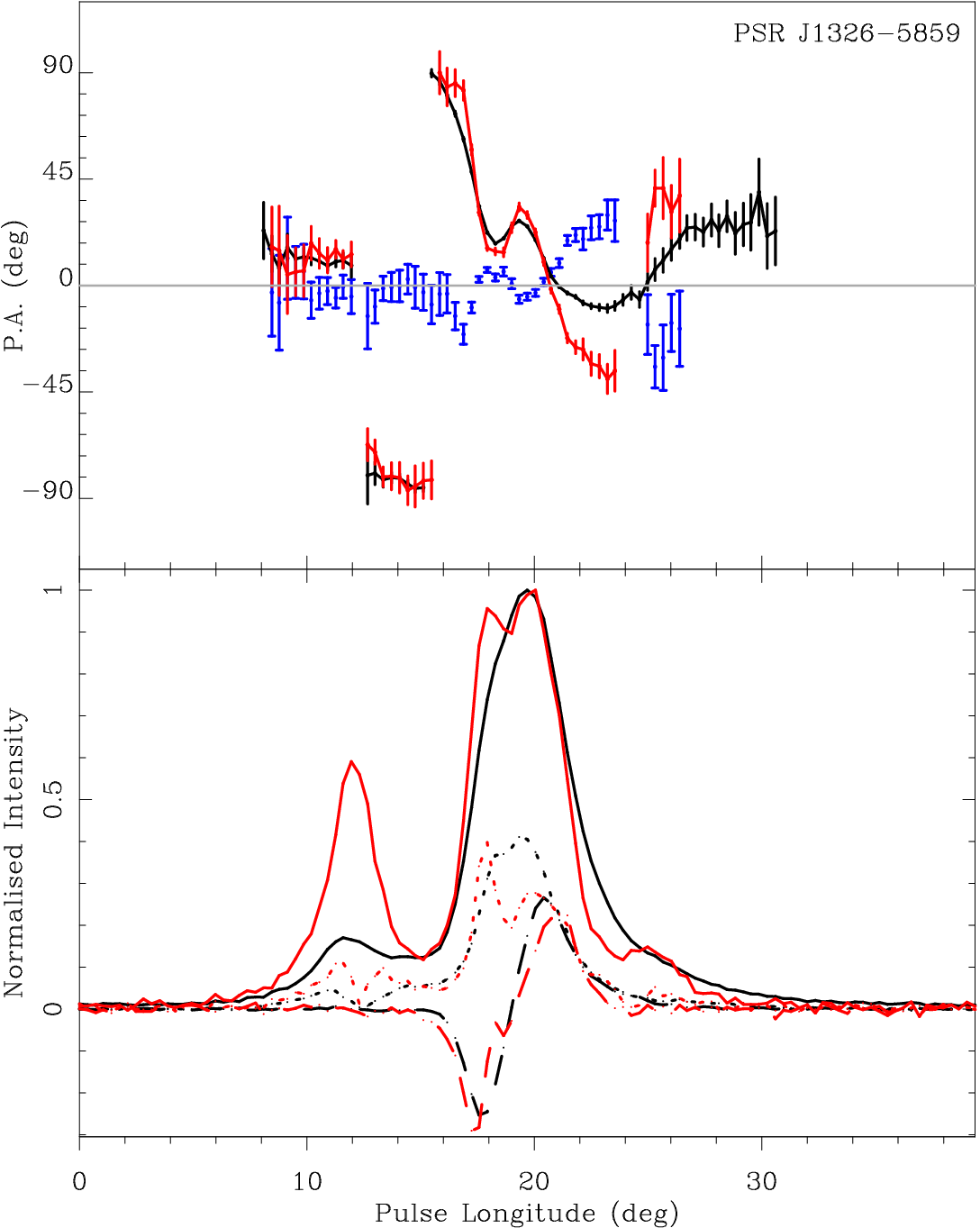}}\\ 
\end{tabular}
\end{center}
\caption{- continued.}
\end{figure*}
\addtocounter{figure}{-1}
\begin{figure*}
\begin{center}
\begin{tabular}{cc}
\resizebox{0.5\hsize}{!}{\includegraphics[angle=0]{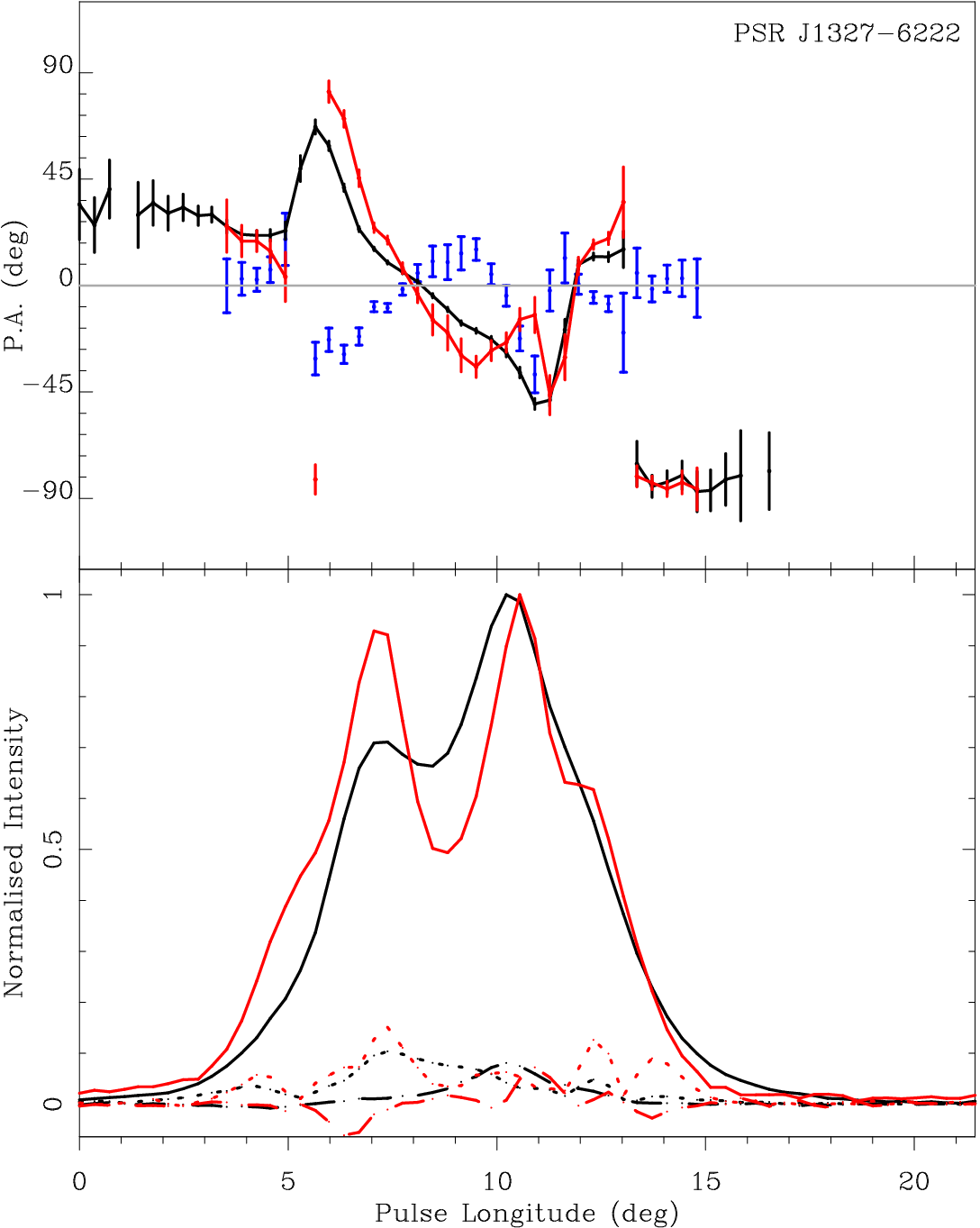}}&
\resizebox{0.5\hsize}{!}{\includegraphics[angle=0]{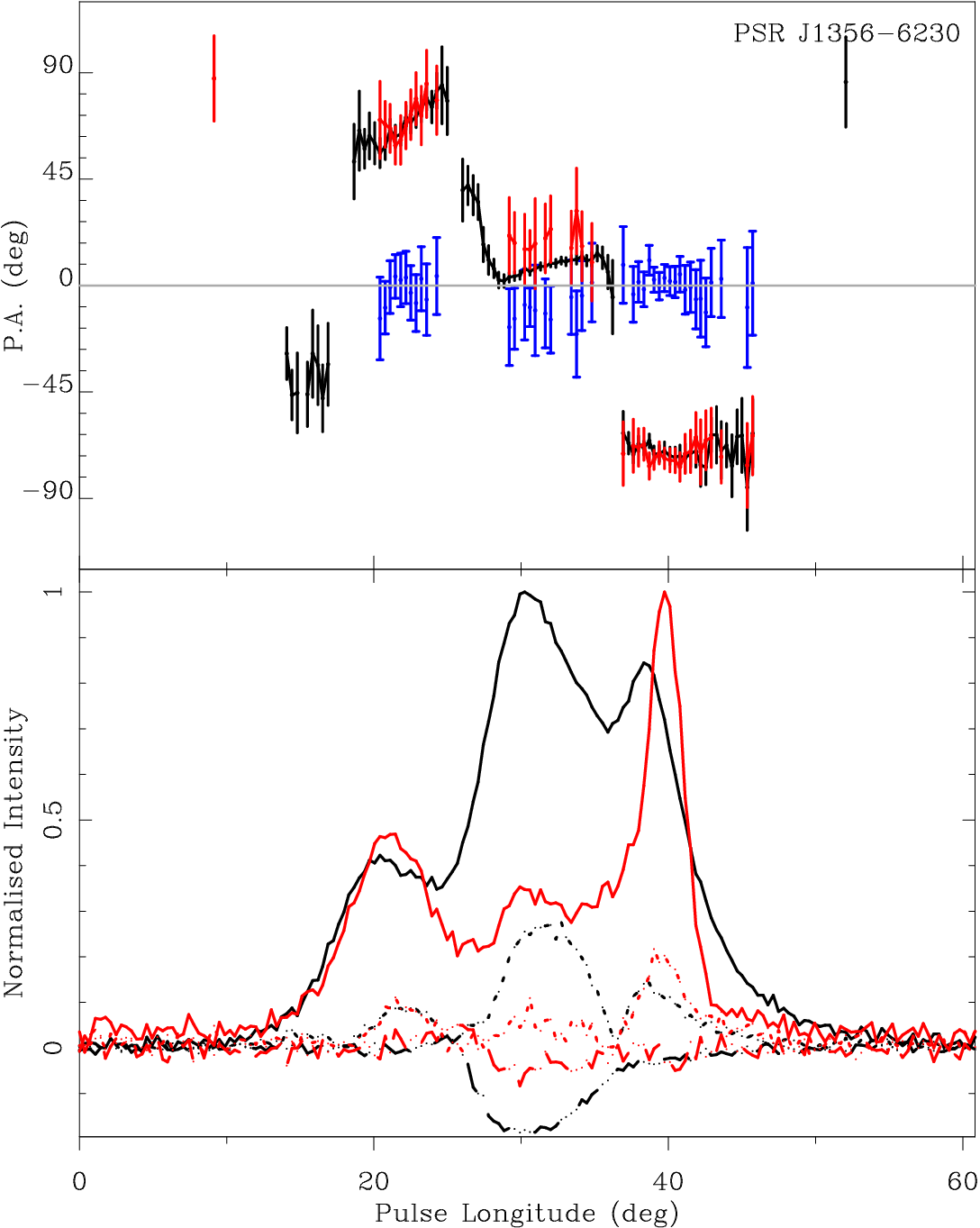}}\\ 
\resizebox{0.5\hsize}{!}{\includegraphics[angle=0]{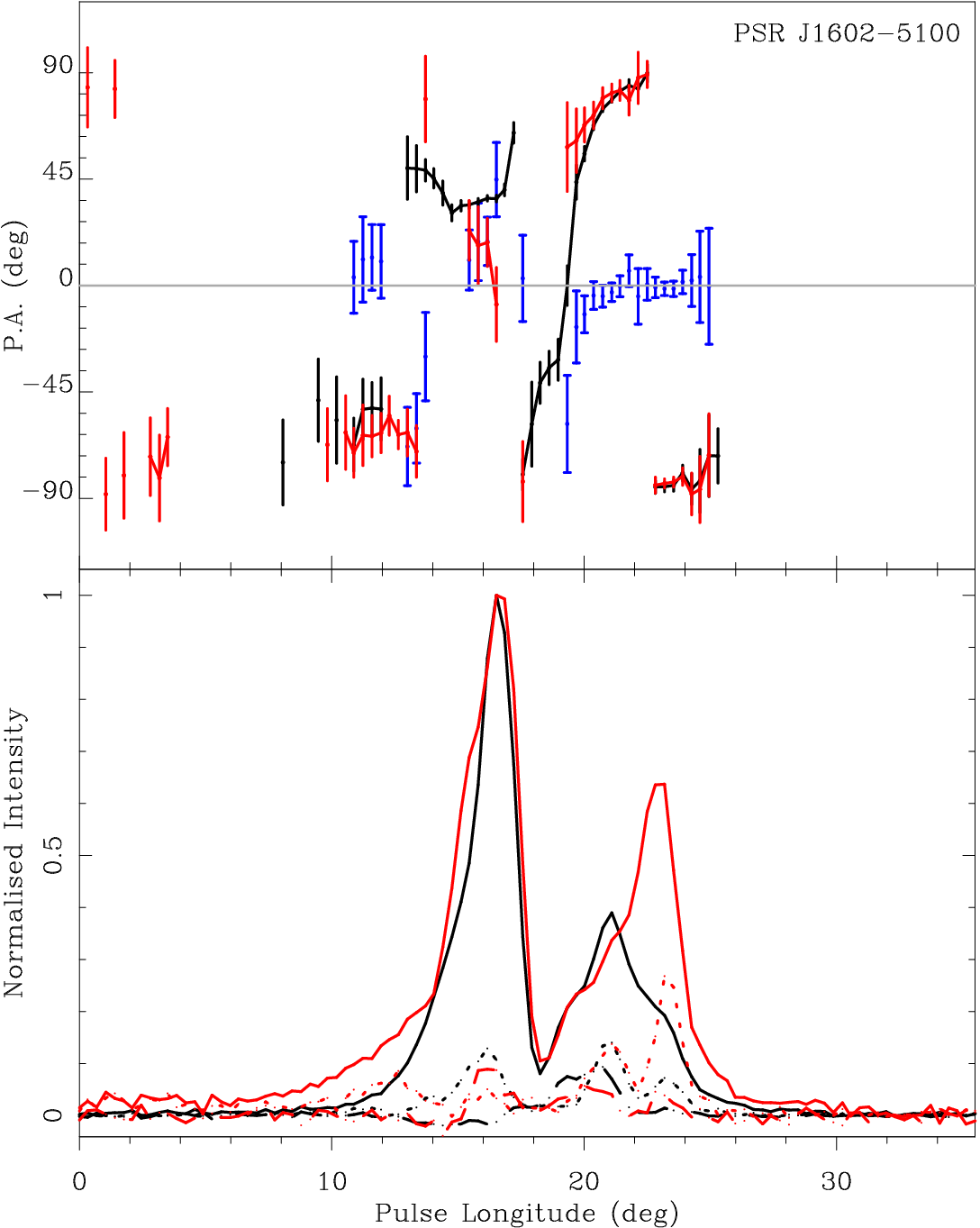}}&
\resizebox{0.5\hsize}{!}{\includegraphics[angle=0]{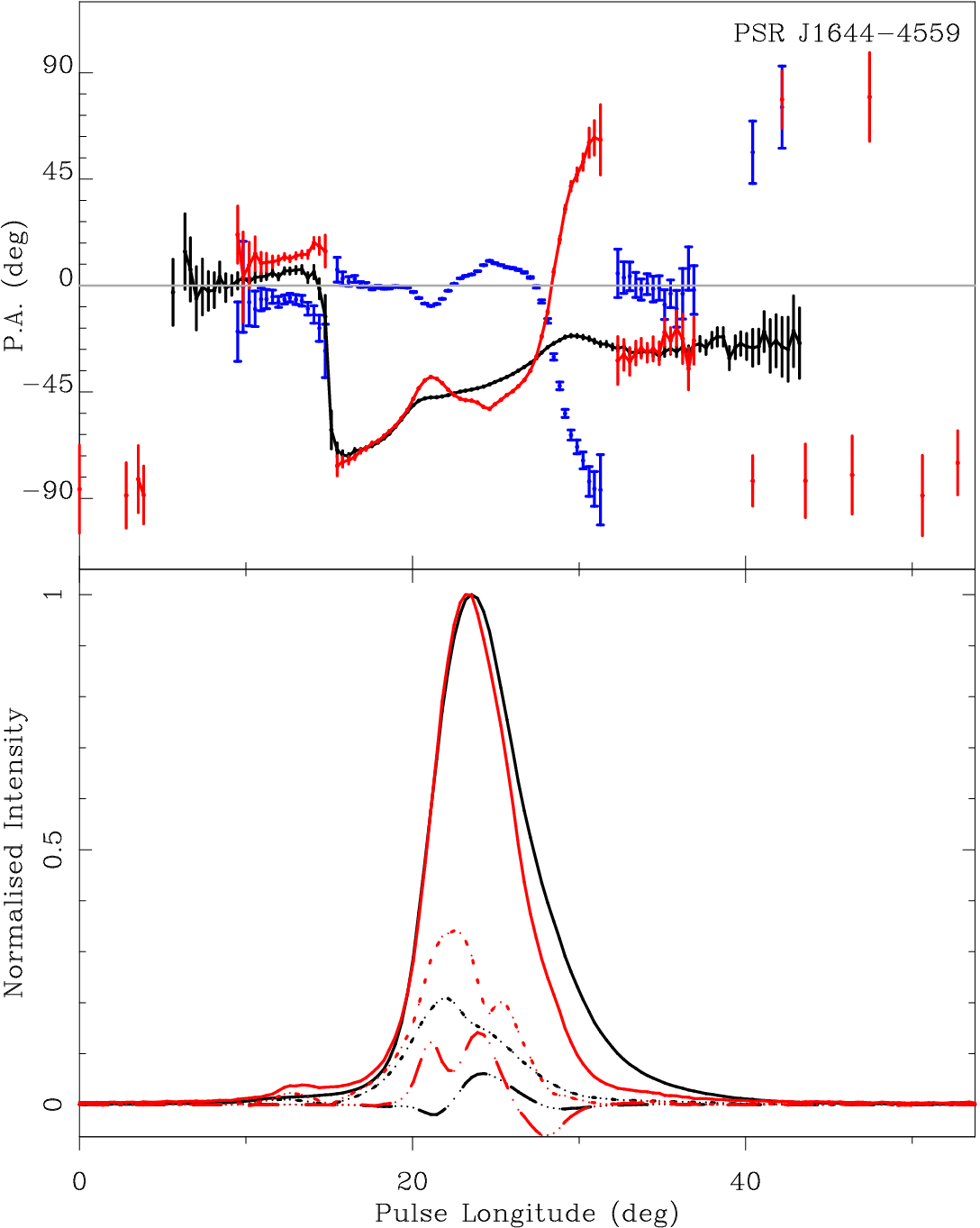}}\\ 
\end{tabular}
\end{center}
\caption{- continued.}
\end{figure*}
\addtocounter{figure}{-1}
\begin{figure*}
\begin{center}
\begin{tabular}{cc}
\resizebox{0.5\hsize}{!}{\includegraphics[angle=0]{J1752-2806.ps}}& 
\resizebox{0.5\hsize}{!}{\includegraphics[angle=0]{J1807-0847.ps}}\\
\resizebox{0.5\hsize}{!}{\includegraphics[angle=0]{J1825-0935.ps}}\\
\end{tabular}
\end{center}
\caption{- continued.}
\end{figure*}
\begin{figure*}
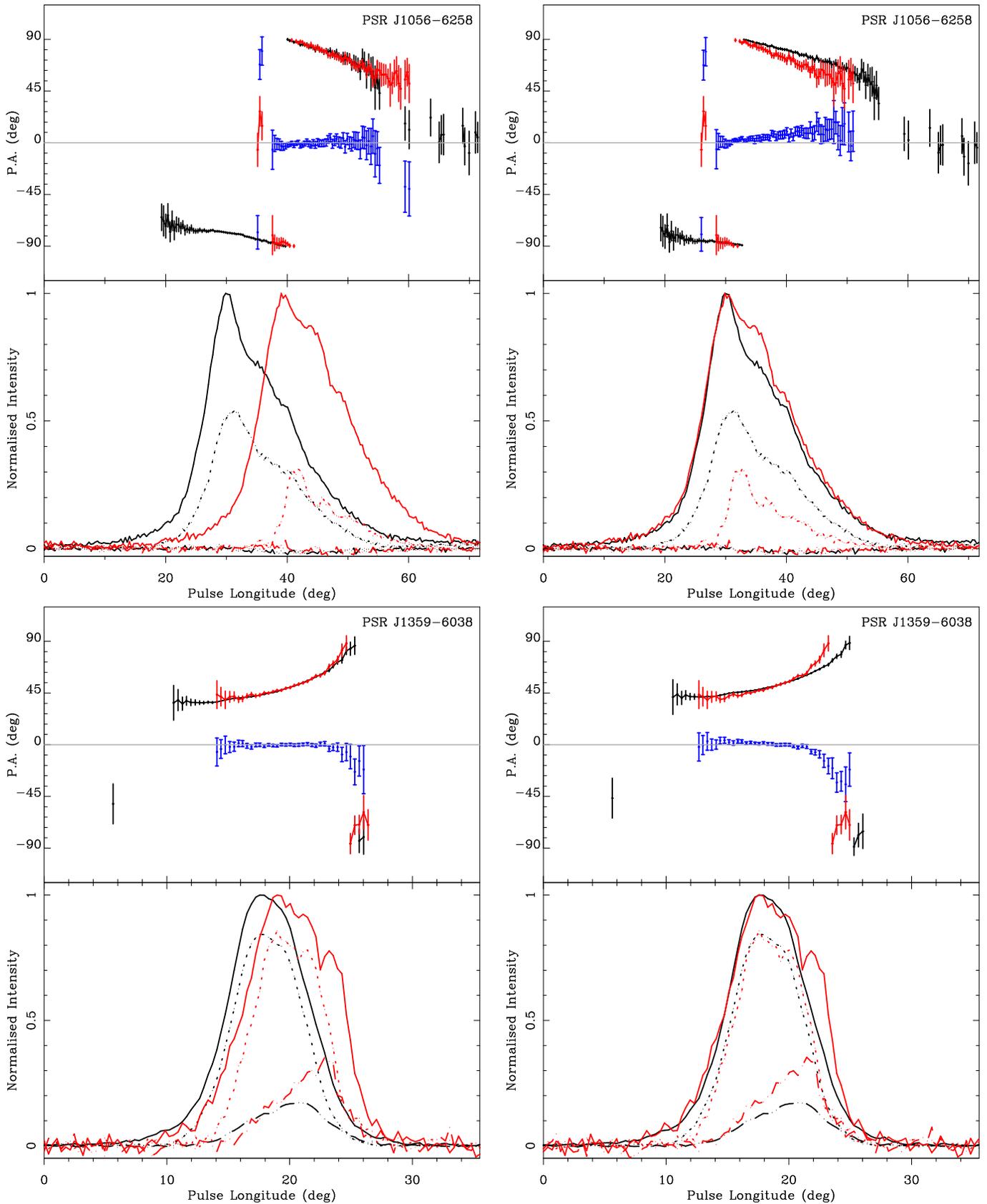

\begin{center}
\begin{tabular}{cc}
\resizebox{0.5\hsize}{!}{\includegraphics[angle=0]{J1056-6258.ps}}&
\resizebox{0.5\hsize}{!}{\includegraphics[angle=0]{J1056-6258b.ps}}\\
\resizebox{0.5\hsize}{!}{\includegraphics[angle=0]{J1359-6038.ps}}&
\resizebox{0.5\hsize}{!}{\includegraphics[angle=0]{J1359-6038b.ps}}\\
\end{tabular}\\
\end{center}
\caption{Same as Figure 1. The top two plots show PSR J1056$-$6258,
  the profile alignment performed on PA (left) and total power
  (right). The bottom two plots show the same for PSR
  J1359$-$6038. The offset between the two alignment methods is 10.7
  ms for J1056$-$6258 and 0.7 ms for J1359$-$6038.}
\end{figure*}

Table 1 shows the list of sources, their rotation period, the pulse
widths at the 10\% level of the peak flux density at both frequencies
and the RM, after removing the ionospheric contribution. For
comparison, we provide the current published RM values where
available.

For an accurate comparison between the polarimetric profiles at 1.375
and 3.1 GHz, we have overlaid them for each pulsar, as shown in Figure
1. Due to the steep flux density spectrum of pulsars, the sources are
much brighter at 1.375 than at 3.1 GHz. To further facilitate
comparisons, we have normalised the flux density at both frequencies
with respect to the peak of the profile. In doing so, the relative
spectral dependence of the various components of the pulse becomes
evident. In most of the pulsars presented here, the strongest profile
component is the same at both frequencies. In these cases, the
normalisation also permits direct comparisons between the fractional
polarisation of these components. We also draw a comparison with
previous data at 400 MHz \citep{hmak77}, 600 MHz \citep{mhma78}, 800
and 950 MHz \citep{vdhm97} and 1612 MHz \citep{mhm80} where
appropriate.

{\bf J0738$-$4042 (B0736$-$40).} The most striking feature of this pulsar is the
smooth PA swing, which is broken twice by orthogonal jumps at the
leading and trailing side of the profile at both 1.375 and 3.1 GHz. In
fact there is little evolution of the polarisation profile between
these frequencies, despite an increase of the total intensity of the
leading component compared to the main peak of the profile at 3.1
GHz. The orthogonal jump in the leading component occurs exactly at
the longitude where both linear and circular polarisation are zero and
is accompanied by a change in the circular polarisation sense at both
frequencies. This is also true for the trailing component jump,
although the change in handedness of the circular polarisation is not
as prominent. Drawing a comparison between our profiles and previous
profiles at other frequencies, we focus on the leading component. The
pulse longitude at which the first PA jump occurs shifts from the
notch between the leading and middle component at 631 MHz to a
slightly earlier longitude at 950 MHz. At that frequency, a second
local minimum of linear polarisation appears at the pulse longitude of
the notch. Surprisingly, the 1612 MHz profile shows the PA in the
region between the local minima in linear polarisation following the
previous swing and the jump occurring at the notch, similar to the 631
MHz data. In the 1.375 GHz profile shown here, the PA agrees with the
950 MHz data. The most plausible picture therefore, involves a change
in dominant OPM between the two local minima of linear polarisation,
occurring between 631 and 950 MHz and subsequently shifting the pulse
longitude of the PA jump at frequencies above 950 MHz.

The comparison of PAs at 1.375 and 3.1 GHz shows that not all three
segments of the PA profile align. Depending on the choice of RM,
either the first or the second and third parts match. We have chosen
the latter in Figure 1, and the differences in PA in the second and
third segments are zero within the uncertainties. Note that the
alignment method results in alignment in the total power of the main
peak despite the misalignment of the linear polarisation.

{\bf J0742$-$2822 (B0740$-$28).} The total power profiles at 1.375 and 3.1 GHz have
the same overall structure. The leading part is made up of two
overlapping components, the middle part of three and the trailing part
of a narrow and a broad component (seven components are also suggested
in \citealt{kra94}). However, the ratios of the component strengths
are different at the two frequencies. In the above order, the first,
fourth and fifth components have flatter spectra than the second and
third components. Arguably, the same is true for the two trailing
components with respect to components two and three. Both profiles are
highly linearly polarised with a small amount, right-hand circular
polarisation across the pulse, in agreement with earlier
observations. The only obvious difference in the PA profiles is a
$90\degr$ jump in the penultimate component at 3.1 GHz, not seen
previously at other frequencies. The best alignment in PA also aligns
the total power.

\citet{gl98} have observed this pulsar in full polarisation at
frequencies up to 1642 MHz, while \citet{hx97} have provided profiles
at 4.85 GHz and 10.55 GHz. The 4.85 GHz profile does not demonstrate
the orthogonal PA jump in the trailing component and the 10.55 GHz
profile lacks a measurement of position angle in this component due to
the low S/N of the linearly polarised flux. Careful investigation of
the frequency dependence of the polarisation in the trailing component
will be useful in determining whether there is a monotonic frequency
behaviour of the strength of each OPM. If so, there should be a
frequency above which the dominant mode switches \citep{kjm05}. Also,
in the trailing component of the 3.1 GHz profile, the mean circular
polarisation is very small across the component. Similar behaviour
(i.e. a drop in the mean linear polarisation, orthogonal PAs and zero
mean circular polarisation) has been observed in a number of pulsars
which show wide distributions of circular polarisation, with many
instances of high left- and high right-hand circular polarisation,
resulting in a considerably higher value for the mean $|V|$
\citep{kjm+03,kj04}. We predict that single pulse observations at high
frequencies will reveal this effect in this pulsar.

{\bf J0835$-$4510 (B0833$-$45).} The profile of the Vela pulsar at 1.375 and 3.1
GHz is made up of three components: two, overlapping at the leading
edge and a weaker one at the trailing edge. The flux density ratio of
these three components with respect to each other is different at the
two different frequencies, in that the second and third components
become more prominent at 3.1 GHz. This pulsar is very highly linearly
polarised at these frequencies, as it is at lower frequencies. The PA
swing of Vela was the original swing used to advocate the RVM.  The PA
profiles at 1.375 and 3.1 GHz are virtually identical. Small
differences arise mainly around the profile centre, where the
aforementioned changes in relative intensity of the overlapping
components also occur. 

{\bf J0837$-$4135 (B0835$-$41).} The profile consists of a central component and
two outriders. The outriders are hardly detectable at lower
frequencies but become more and more prominent at 1.375 and 3.1
GHz. The PA swing in the outriders is similar at both frequencies,
with a difference of zero within the errors. In the middle component,
there are two peaks of linear polarisation at both frequencies, with
different ratios. The largest disagreement in PA also occurs within
this range of the profile. An abrupt PA discontinuity at 1.375 GHz can
be seen between the leading and middle component, despite a much
smoother transition at 3.1 GHz. The trailing peak of linear
polarisation in the middle component shows a flat, identical PA at
both frequencies. Also, there is less overall circular polarisation in
the middle component of the 3.1 GHz profile with respect to 1.375
GHz. The jump in the 1.375 GHz profile is also seen at lower
frequencies.

{\bf J0942$-$5552 (B0940$-$55).}  The three main components of this pulsar show the
archetypal behaviour of outriders coming up with respect to the middle
component at higher frequencies. The PAs at 1.375 and 3.1 GHz are
identical across the pulse profile. The central component at 3.1 GHz
seems to lag its 1.375 GHz counterpart. The profile of linear
polarisation in the middle component aligns well, despite an extra
peak at 3.1 GHz that does not change the overall width of the
component.

{\bf J1056$-$6258 (B1054$-$62).} The profiles of this pulsar at 1.375 and 3.1 GHz
are shown in the top row of Figure 2. The graph on the left shows the
alignment preferred by the PA matching routine, which results in an
obvious offset between the total intensity profiles. On the right, the
total intensities have been matched. The offset between the two cases
is 10.7 ms. Using the model of \citet{bcw91}, it is possible to
convert this delay into a difference in an emission height, without
obtaining the absolute emission height at each frequency, according to
the simple equation \citep{ml04}
\begin{equation}
\Delta R \approx -\frac{c}{4}\times \Delta t
\label{delay}
\end{equation}
where $c$ is the speed of light. For this pulsar, we obtain an
emission height difference of $\sim 800$ km, which is unlikely.


Another possible cause of the alignment discrepancy is that the
overlapping linearly polarised components change their relative
intensity at 3.1 compared to 1.375 GHz, thereby intrinsically changing
the PA profile. The change is such, that the best PA alignment results
in the offset in the total power profile.  This explanation then
suggests that the total power profiles aligned between the two
frequencies is correct and not the PA alignment. The RM computed is
then different for the two alignment methods; using the PA alignment,
the RM is $-1.0$~rad~m$^{-2}$, where as the total power alignment
results in an RM of 6.0 rad~m$^{-2}$, which is much closer to the
previous published value of 4.0 rad~m$^{-2}$ \citep{cmh91}.

Comparing the profiles of this pulsar, reveals a large reduction in
the linear polarisation at 3.1 GHz, especially in the leading part of
the profile. A $90\degr$ jump in PA is seen at 3.1 GHz lending support
to the hypothesis that the depolarisation is due to competing OPM
\citep{kjm05}. Also, there is almost no integrated circular
polarisation at either frequency.

{\bf J1157$-$6224 (B1154$-$62).}  At 3.1 GHz there are prominent outriders in the
total power profile compared to the 1.375 GHz profile. The linear
polarisation in the middle component is marginally less at 3.1 GHz
than at 1.375 GHz. The PA swing of the both profiles is complex but
identical for the pulse longitudes where the linear polarisation is
high. The circular polarisation shows an S-shaped swing around the
centre of the pulse, often associated with emission originating from
close to the magnetic axis \citep{ran86}, despite the flat PA
profile. At low frequencies this pulsar has a single-component profile
with some linear polarisation and a flat PA profile.

{\bf J1243$-$6423 (B1240$-$64).} There is more outlying total
power emission on either side of the central components of this pulsar
at 3.1 than at 1.375 GHz. A decrease in linear polarisation with
frequency is also evident, despite both profiles having fairly large,
right-hand circular polarisation. The PA profile is characterised by
two discontinuities: the first, at pulse longitude 12\degr\, appears
to be orthogonal at 1.375 GHz and only $\sim 60\degr$ at 3.1 GHz; the
second, just before pulse longitude 20\degr\, occurs at a very steep
part of the PA swing and is not orthogonal at either frequency. The
main differences in the absolute PAs between 1.375 and 3.1 GHz are
concentrated in a window between pulse longitudes 16\degr\ and 20\degr\, which
coincide with the steepest parts of both PA curves. It is interesting
that the general shape of the PA profile at both frequencies resembles
a $\upsilon$, with a flat segment on either side, allowing for $90\degr$
discontinuities. In previous observations the jump at the trailing
edge is seen for the first time at 1612 MHz, whereas it is not seen at
950 MHz or 631 MHz.
 

{\bf J1326$-$5859 (B1323$-$58).} At 1.375 GHz, the profile of this pulsar consists of
a weak component, followed by a strong middle component which trails
off, hinting at the presence of a weak trailing component. The linear
polarisation, which is moderately high across most of the pulse, shows
a minimum under the leading component, coinciding with a $90\degr$
discontinuity in the PA. There is a swing in the circular
polarisation, from right to left hand, in the centre of the pulse. At
3.1 GHz, the total power profile looks more complicated. The PA
profile resembles the lower frequency profile, with some notable
differences. There is an orthogonal PA jump at the leading part of the
profile, as at 1.375 GHz. Both PA profiles feature kinks and bends,
particularly under the middle total power component. The linear
polarisation there consists of two overlapping components, more
clearly seen at 3.1 than at 1.375 GHz and the PA swing is also more
perturbed at the higher frequency. There is also a small right-hand
circular polarisation feature just before the change in handedness in
the 3.1 GHz profile. Also, the smooth positive slope in PA joining the
middle and trailing parts of the 1.375 GHz profile becomes what appears
to be an orthogonal PA transition after a minimum in linear
polarisation at 3.1 GHz. The total power components have different
evolution with frequency, the leading component having the flattest
spectrum. The second component of the central pair, which has a
steeper spectrum with respect to its neighbours, also seems to
de-polarise significantly.

 {\bf J1327$-$6222 (B1323$-$62).} The profile of this pulsar consists of at least
four components, which are more distinguishable at 3.1 than at 1.375
GHz. There is little linear and circular polarisation in both
profiles. The PA profile at the two frequencies does not resemble the
S-shaped curve of the RVM. In fact, between pulse longitudes 5\degr\
and 13\degr\, the PA profiles are $\upsilon$ shaped. At pulse
longitude 13\degr\ an orthogonal jump occurs at both frequencies,
whereas at longitude 5\degr\ the steep increasing PA at 1.375 GHz
becomes an apparent $90\degr$ discontinuity at 3.1 GHz. The PAs at
1.375 and 3.1 GHz agree over approximately half the profile, namely
the leading and trailing ends. Significant deviations can be found in
the middle part of the profile.

Out of the two strongest components, the leading one has a
comparatively flatter spectrum and more linear polarisation at 3.1
than at 1.375 GHz. The trailing edge of the profile also appears more
fractionally polarised at 3.1 than at 1.375 GHz.

{\bf J1356$-$6230 (B1356$-$60).} At 1.375 GHz, the profile of this pulsar consists of
a strong middle component flanked by outriders. Linear polarisation is
moderate and a large amount of right-hand circular polarisation is
seen in the middle of the pulse. At 3.1 GHz, the outriders are
stronger than the middle component, and the degree of polarisation is
higher in most of the pulse. The circular polarisation component seen
at 1.375 GHz is not present at 3.1 GHz.  As with J0738$-$4042, the PA
profile at both frequencies consists of three segments, only two of
which align between 1.375 and 3.1 GHz. The PA under each of the three
components does not vary, with a similar value in the two outer
components, displaced with respect to the flat PA in the middle
component.

{\bf J1359$-$6038 (B1356$-$60).} This pulsar is shown in the second row of Figure
2. Similar to PSR J1056$-$6258, the total intensity profile at 3.1 GHz
lags the 1.375 GHz profile when attempting an alignment based on PA,
as shown in the left plot. To align the total intensity profiles as
shown in the right plot, a shift of 0.74 ms is required. Equation
\ref{delay} translates this delay into to $\sim 55$ km of difference
in emission altitude. However, the additional component coming up at
3.1 GHz at the trailing edge of the pulse may be responsible for the
PA differences observed there when aligning the total power.  This
pulsar is highly linearly and circularly polarised, with increasing
circular polarisation at 3.1 GHz. The PA profiles also look identical
between the two frequencies.

{\bf J1602$-$5100 (B1558$-$50).} At 1.375 GHz, this pulsar has two main peaks made
up of multiple components. The PA profile shows two flat segments,
joined in the middle by a very steep swing, that appears to
continuously sweep over $225\degr$.  There is an orthogonal PA jump at
1.375 GHz at the very leading edge of the profile (pulse longitude of
$12\degr$). At 3.1 GHz, the last component of the trailing peak
becomes more prominent in comparison with the other components, as
does the linear polarisation in the same component. Despite the large
complexity of the PA profile at 1.375 and the few significant PA
measurements at 3.1 GHz, the PAs mostly match well, especially at the
leading and trailing edges of the pulse.

{\bf J1644$-$4559 (B1641$-$45).} The PA profiles of this pulsar resemble the
$\upsilon$ shaped profiles of PSR J1327$-$6222. This pulsar is
moderately linearly polarised, the fractional linear polarisation
increasing with frequency. There is an orthogonal PA jump at the
leading edge of the profile, seen at both 1.375 and 3.1 GHz, whereas
an additional $90\degr$ PA jump is seen at the trailing edge of the
3.1 GHz profile, proceeded by a non-instantaneous transition of $\sim
90\degr$. The leading and trailing parts of the PA profile are flat
and do not simultaneously align. In the middle of the profile,
although the two PA curves have the same overall swing, the kinks are
more pronounced at 3.1 than at 1.375 GHz. There is also a change in
the sense of the circular polarisation around pulse longitude 26\degr\
. Previous observations at lower frequencies show mainly the effects
of scattering on the profile.

{\bf J1752$-$2806 (B1749$-$28).} At both frequencies, the profile of this pulsar
shows little linear polarisation and circular polarisation that swings
from right to left hand. The best alignment of PAs between the two
frequencies occurs at the leading part of the profile, where the PA
has a steep negative slope. However, the PAs disagree after pulse
longitude 10\degr\ . This coincides with the part of the pulse that
has a steeper spectral index. An orthogonal PA jump near the leading
edge of the profile is observed at 631 MHz and 950 MHz in previous
profiles, but not at 400 MHz and 1612 MHz, although there is very
little linear polarisation to make a PA measurement. From the 1.375
GHz profile shown here we suspect that the steep negative PA slope at
the leading part of the profile can give the impression of an
orthogonal jump in observations with poor temporal resolution.

{\bf J1807$-$0847 (B1804$-$08).} This is another case of a profile consisting of a
central component with outriders coming up at 3.1 GHz. Low to moderate
linear and circular polarisation are seen across the profiles. The
minimum in linear polarisation in the leading component at both
frequencies, occurs at the same time as an orthogonal PA jump. At
1.375 GHz there is a second orthogonal PA jump in the trailing edge of
the profile, which can also be inferred from the 3.1 GHz profile. A
systematic deviation in PA between the 2 frequencies occurs in the
middle part of the profile. At 950 MHz, $90\degr$ PA jumps in the leading
and trailing edge of the profile can be seen, much like the profiles
shown here.

{\bf J1825$-$0935 (B1822$-$09).} As is the case with the previous pulsar, there is
little change in the polarisation profile between 1.375 GHz and 3.1
GHz (other polarisation profiles can be found in \citealt{gl98},
\citealt{hx97}). The leading component remains very highly linearly
polarised and the kinks in the PA profile of the trailing component
coincide with a dip in the linear polarisation.


\begin{figure*}
\resizebox{\hsize}{!}{\includegraphics[angle=-90]{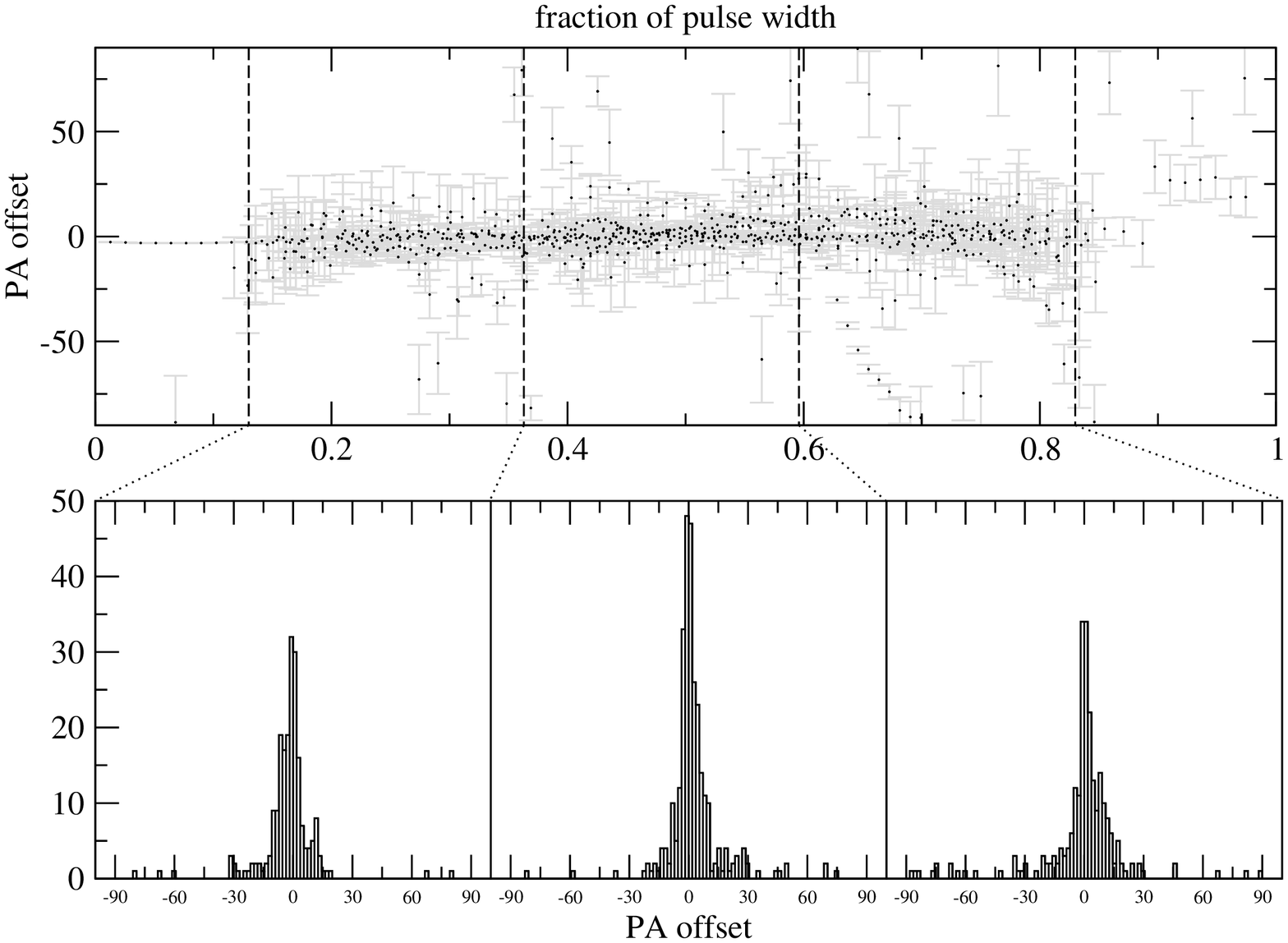}}
\caption{Top panel: The difference in PA between 1.375 and 3.1 GHz, as
  a function of the position within the width of the pulse, where PA
  measurements are possible.}
\end{figure*}

\section{Discussion}
Of the pulsars studied here, only Vela, PSRs J0742-2822, J1056-6258
and J1359-6038 show a smooth, unbroken PA swing expected under the
RVM.  RVM fitting can be performed with confidence only in the former
two cases. For 10 of the pulsars in our list, $\alpha$ has been
derived in \citet{ran93, ran93b}, by using the width of the central
component. The derived values are mostly in agreement with geometrical
values presented by \citet{lm88}, who instead made use of the
RVM. However, as is typical of the pulsar population generally
(especially at high frequencies), computing geometrical parameters
such as $\alpha$ and $\beta$ must necessarily be subject to large
uncertainties.  We therefore do not concentrate on the geometrical
implications of these results; rather we focus on possible causes of
PA variations and their frequency dependence based purely on a direct
comparison between the PA curves. We have made note of observations at
other frequencies but caution that observations made with low time
resolution can often be misleading.

We do note however, that many of the pulsars in our sample show
complex (multi-component) profiles and large overall PA swings. This
likely indicates that $\beta$ is small in these pulsars and the
sightline traverses close to the magnetic pole.  Also, deviations of
the PA from smooth RVM curves tend to occur mostly in the centre of
profiles either because components overlap there or perhaps for some
physical reason assoicated with core emission.  Observations at even
higher frequencies, where individual components are narrower and the
core emission much less prominent would then reveal smoother PA
curves, an effect which is seen, at least in some pulsars
\citep{hkk98}.


The 17 pulsars presented in Figures 1 and 2 and described in the
previous section demonstrate the diversity in emission properties.
The number of total intensity components, the degree of polarisation
and the swing of the PA with pulse longitude vary significantly from
pulsar to pulsar. Superposing the pulse profiles at 1.375 and 3.1 GHz
reveals differences in the emission properties between these two
frequencies. These differences, noted in the previous section, mostly
relate to the ratios of the various component amplitudes within a
profile and the varying degree of linear and circular polarisation of
individual components at the two frequencies. The changing component
amplitude ratios are a manifestation of the different spectral
behaviour of different components, a topic extensively discussed
recently by \citet{kjm05}.

The pulse widths presented in Table 1 reveal little change between
1.375 and 3.1 GHz. This is not surprising, considering that both these
frequencies are relatively high. The flaring of the last open magnetic
field lines that results in broader profiles at lower frequencies does
not appear to play a significant role here. \citet{mr02a} also point
out that components that originate from magnetic field lines closer to
the magnetic axis have the same widths across wider frequency
ranges. It is therefore evident that information on the difference in
emission heights cannot be gleaned from the pulse widths. For the
pulsars of Figure 2, namely J1056$-$6258 and J1359$-$6038, the PA
alignment method may not be accurate, due to intrinsic PA changes. For
the remaining pulsars we observed, the PA method aligns the profiles
to $\pm 1$ data bin, the sampling time of which is equal to the pulse
period over 1024. This places a limit on the range of emission height
differences between 1.375 and 3.1 GHz for our pulsars, ranging from
$\sim 6.7$ km for the Vela pulsar to $\sim 64.8$ km for J1602$-$5100,
as computed from Equation \ref{delay}.

There are also 4 pulsars in our source list (PSRs~0942$-$5552,
1157$-$6224, 1356$-$6230 and 1807$-$0847) that have profiles with an
outer component pair flanking a central component. In all 4, the
relative intensity of the outer components with respect to the central
component is higher at 3.1 than at 1.375 GHz. However, none of these
sources show spectacular effects when it comes to the PA evolution
with frequency. In fact, the first three pulsars show flat PAs across
the central component at both frequencies and only PSR J1807$-$0847
has a steep PA swing in the middle of the profile. Unfortunately,
however, the linear polarisation is very low in the central component
of this pulsar, making the frequency comparison close to impossible.

For these 4 pulsars it is possible to derive absolute emission heights
at each frequency, just by measuring the offset of the central peak
from the middle of the outer components \citep{gg01,drh04}. Although
the direction of the offset is in agreement with the fact that the
central component is emitted at a lower height (the leading outer
component is always closer to the central component than its trailing
counterpart), it is clear that the offset is identical at the two
frequencies for all 4 of the aforementioned pulsars. This constitutes
further evidence that the emission heights are very similar between
the two observed frequencies. The absolute emission heights derived,
using equation 7 of \citet{drh04}, are 14.5, 80, 33.5 and 4.5 km
respectively for the aforementioned 4 pulsars. PSR 1807$-$6230 has the
shortest rotation period from this selection, limiting the radius of
its magnetosphere.

In the PA profiles, there is an obvious overall similarity between the
two frequencies. Occasionally, the PA is different in narrow or
extended parts of the profile, although even then, the agreement
between frequencies outweighs the disagreement. To illustrate this, we
have summarised in Figure 3 the PA differences $d$PA in all the
pulsars we observed. We initially identified the pulse widths over
which PA measurements were possible, for all the pulsars. The
abscissae on the top panel denote the pulse longitude as a fraction of
this width. Note that these widths are sometimes greater than the
W$_{10}$ widths of Table 1. This creates a normalised pulse longitude
scale for all the pulsars. The points on the plot represent the
difference in PA, or $d$PA, as shown in the plots of Figures 1 and 2,
plotted against their location within the pulse.  Furthermore, the
second panel in Figure 3 shows the histograms of $d$PA in each of
three regions of the top panel. The intention is to demonstrate that
the distribution of $d$PA does not obviously depend on the location
within the pulse. The standard deviation of all the $d$PA values in
the top panel is $16.6\degr$, compared to $13.9\degr$, $13.8\degr$ and
$21.2\degr$ in the three regions of the bottom panel. The higher
standard deviation value of the third region is partially caused by
the large negative $d$PA values in PSR J1644$-$4559. The average error
in $d$PA is $\sim 7$, so the standard deviations measured are between
2 and 3 times this value, which consolidates the observation that the
PA differences cannot merely be put down to noise.

The histograms in the bottom row of Figure 3 can be used to identify
OPM as a possible source of PA disagreement between 1.375 and 3.1
GHz. If at some intermediate frequency, the dominant mode of
polarisation has changed, the PA at the high frequency will be
orthogonal to the low frequency. Interestingly, in Figure 3 there
appears to be a Gaussian spread of $d$PA around its mean of zero. The
instances of $d$PA close to $90\degr$ or $-90\degr$ are few and do not
represent a clear sub-group of the $d$PA distributions. This fact
suggests that for our sample of sources, changes in the dominant OPM
are not common between these two frequencies. The general decrease in
the linear polarisation, however, is an indication that the strengths
of the OPMs are becoming more equal \citep{kkj+02}.

The measured standard deviation in the distribution of $d$PA places
constraints on the frequency dependence of possible mechanisms that
set the PA. If the superposed modes of polarisation are not orthogonal
and have different spectral behaviour, a dependence of the PA on
frequency will arise. The different spectral behaviour is evident by
the change in linear polarisation between the two frequencies. The
left plot in Figure 4 shows the effect of superposed, non-orthogonal
modes of polarisation on the observed degree of linear polarisation
and PA. From left to right, the lines correspond to a deviation from
orthogonality of $1\degr$, $3\degr$, $5\degr$, $10\degr$ and
$20\degr$. The vertical line has a length of $13\degr$, as a measure
of the standard deviation of $d$PA. This figure demonstrates how large
$d$PAs combined with small changes in the degree of linear
polarisation, require larger departures from orthogonality of the
polarisation modes. It also demonstrates that non-orthogonality of the
modes results in the minimum degree of linear polarisation not being
zero, as would be the case with orthogonal modes. More importantly
however, this figure shows that in cases where the degree of linear
polarisation changes significantly from one to the other frequency,
non-orthogonality will lead to larger values of $d$PA. We conducted
the following test: for all bins where the fractional linear
polarisation is greater than 0.5 at 1.375 GHz, we measure the
difference of the degree of linear polarisation and $d$PA between the
two frequencies. The results are shown in the right plot of Figure 4.
There is a hint that the points which correspond to a large change in
the fractional linear polarisation, also correspond to larger
$d$PA. The correlation however is very weak, which demonstrates that
an explanation involving non-orthogonal polarisation modes is
unlikely.

In the individual plots of Figures 1 and 2, we identify two types of
deviation of $d$PA from zero. The first consists of a part of the
profile where the PAs at the two different frequencies deviate in a
quasi-random way, resulting in a variable $d$PA which gradually
returns to zero. Good examples of this behaviour are PSRs J1243$-$6423
and J1327$-$6222. In the second type $d$PA has a constant but non-zero
value across a fraction of the pulse profile. The PAs at 1.375 and 3.1
GHz run parallel to each other and then converge to being
identical. Examples of this can be found in PSRs J0738$-$4042 and
J1644$-$4559.

In the pulsars that belong to the first type, the total power profiles
are complex with generally low linear polarisation. We can often
discern what appear to be several, partially overlapping components of
linear polarisation, coinciding with the most perturbed parts of the
PA swing. The question then arises whether the observed changes in PA
are merely due to changes in the relative strengths of overlapping
components of linear polarisation. It seems appropriate to conduct
such tests for PSRs J1243$-$6423 and J1327$-$6222, to formulate a
model. From our data, the ratios of the various components in e.g. PSR
J1243$-$6423 are certainly different at 1.375 and 3.1 GHz, however
single pulse data is required to properly account for the effects of
superposed modes of polarisation \citep{kjk03}. Also, in the case of
the Vela pulsar where the linear polarisation is very high, the second
component in the leading edge of the profile becomes more prominent at
3.1 GHz and at the same time the PA profile is distorted at the same
pulse longitudes. An interpretation based on overlapping components
with intrinsically different PAs again seems likely.

For the second type of $d$PA offsets, the same reasoning cannot
explain the observations. It is tempting to attribute the fact that
some components show a constant offset in PA between 1.375 and 3.1
GHz, e.g. the leading component of PSR J0738$-$4042, to a different
Faraday Rotation measure. Although we cannot rule this possibility
out, there seems to be little support in terms of the underlying
physics. However, a proper test requires one more observing frequency,
to check if the dependence of PA on frequency is of the same nature as
Faraday Rotation. The fact that the observed difference in PA
corresponds only to a few units of rotation measure does not permit us
to conduct this test within one of the bands of observation.

\begin{figure*}
\centerline{
\resizebox{0.45\hsize}{!}{\includegraphics[angle=-90]{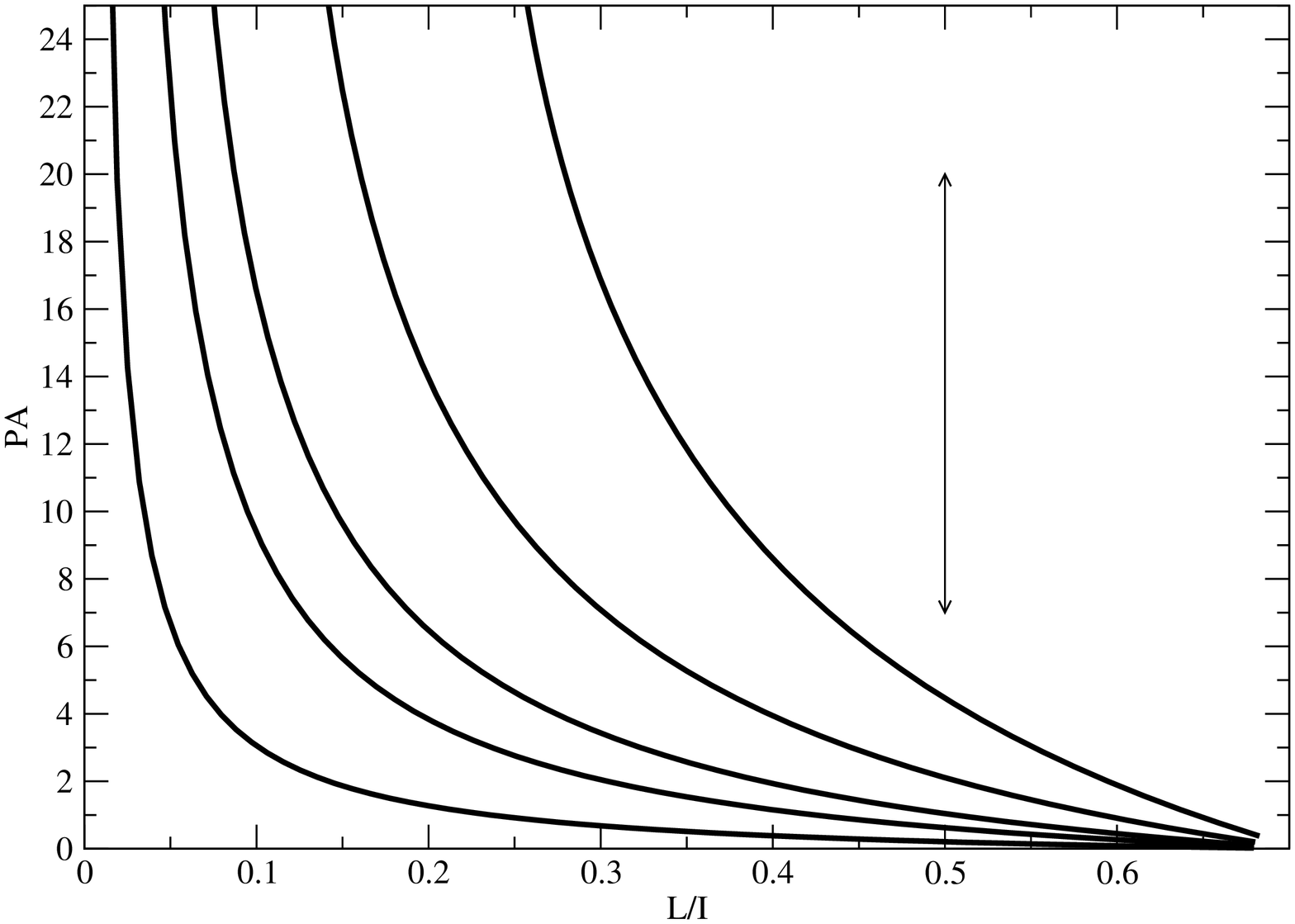}}
\resizebox{0.45\hsize}{!}{\includegraphics[angle=-90]{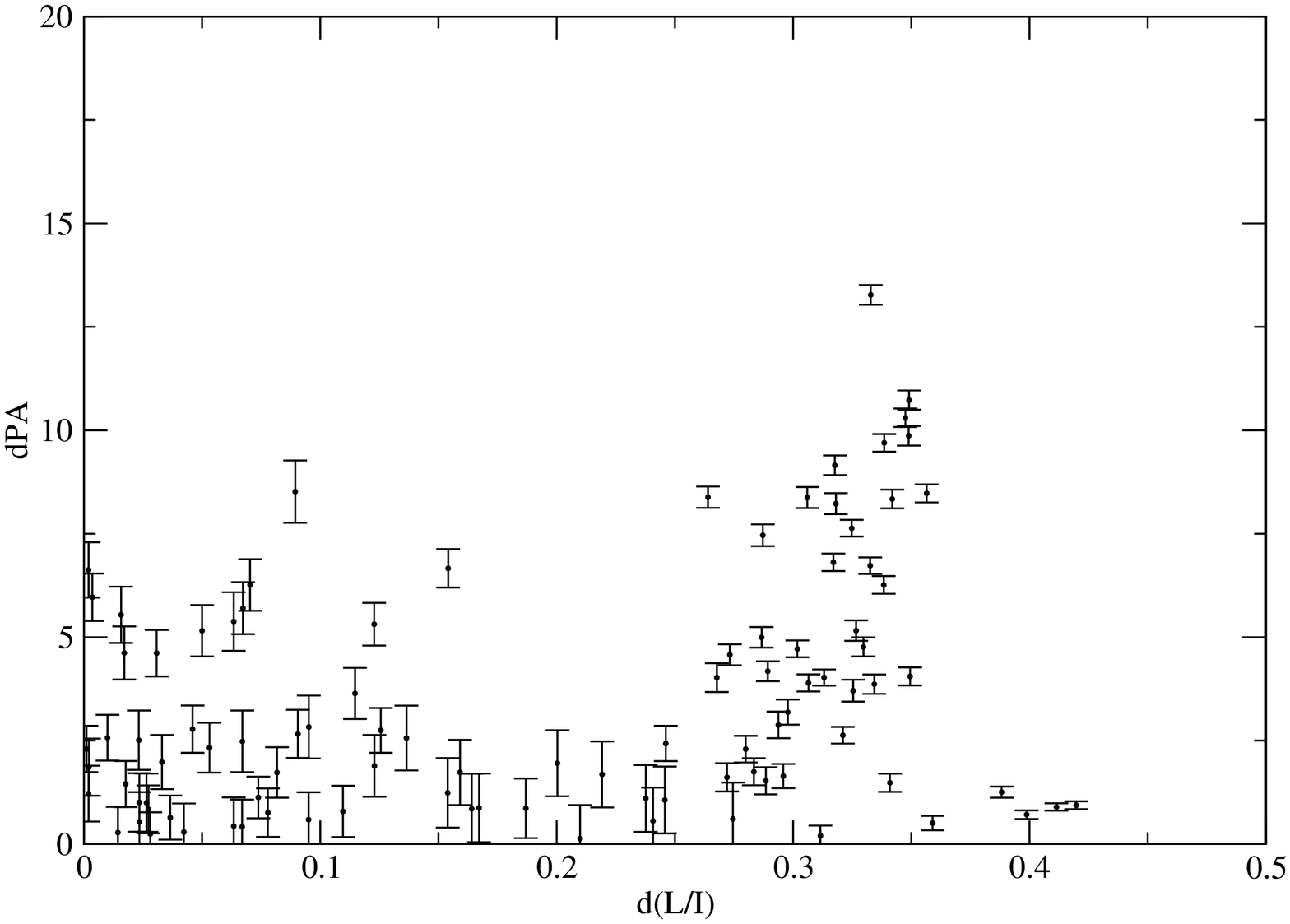}}}
\caption{Left: The effect of superposed, non-orthogonal polarisation
  modes on the observed PA and degree of linear polarisation
  (L/I). The different curves, from left to right, correspond to a
  deviation from orthogonality of $1\degr$, $3\degr$, $5\degr$,
  $10\degr$ and $20\degr$. Right: $|d$PA$|$ plotted against the
  difference in the fractional linear polarisation, for all instances
  of high linear polarisation at 1.375 GHz.}
\end{figure*}
\section{Conclusions}
We have presented for the first time superposed polarimetric profiles
of 17 pulsars at 1.375 and 3.1 GHz, with absolute values of PA. The
presentation demonstrates the frequency dependence of the PA as a
function of pulse longitude. We find that, in general:
\begin{itemize}
\item The PA profiles are similar at the two frequencies, despite the
lack of resemblance to the RVM curve. 
\item We can identify a tentative association between the changes in
the PA between 1.375 and 3.1 GHz and the change in relative amplitude
of overlapping total power components. This observation clashes with
the RVM in that the polarization of the emission at each pulse
longitude is constrained by the pulse component in which it belongs,
rather than being determined purely by geometry.
\item The similarities permit an alignment method based on the PA
profile, which allows us to calculate very accurate RMs. The method
performs very well for all but two pulsars: J1056$-$6258 and
J1359$-$6038, where an explanation related to intrinsic development of
the PA profile seems more likely. 
\item In our sample, there are not many instances of the PA at one
frequency being orthogonal to the other. 
\item Using our measurements of the fractional linear polarisation, we
place constraints on the degree of non-orthogonality between the
polarisation modes.
\item Although in most pulsars the differences in PA between the two
frequencies vary on short scales of pulse longitude, in some pulsars
such as J0738$-$4042, we have found a constant offset in PA across
extended parts of the profile. We intend to perform specific
observations on this pulsar to identify whether this offset can be
accounted for by a different RM value, or some other physical effect.
\end{itemize}

\section*{Acknowledgments}
The Australia Telescope is funded by the Commonwealth of Australia for
operation as a National Facility managed by the CSIRO.  We thank
J.~Han, J.~Weisberg and R.~Manchester for providing us with
ionospheric RM calculation algorithms and R.~Edwards for useful
suggestions.
\label{lastpage}
\bibliographystyle{mn2e}
\bibliography{journals,modrefs,psrrefs,crossrefs}

\begin{thebibliography}{}

\bibitem[\protect\citeauthoryear{Backer \& Rankin}{Backer \&
  Rankin}{1980}]{br80}
Backer D.~C.,  Rankin J.~M.,  1980, ApJS, 42, 143

\bibitem[\protect\citeauthoryear{Backer, Rankin \& Campbell}{Backer
  et~al.}{1975}]{brc75}
Backer D.~C.,  Rankin J.~M.,    Campbell D.~B.,  1975, ApJ, 197, 481

\bibitem[\protect\citeauthoryear{Blaskiewicz, Cordes \& Wasserman}{Blaskiewicz
  et~al.}{1991}]{bcw91}
Blaskiewicz M.,  Cordes J.~M.,    Wasserman I.,  1991, ApJ, 370, 643

\bibitem[\protect\citeauthoryear{Cordes}{Cordes}{1978}]{cor78}
Cordes J.~M.,  1978, ApJ, 222, 1006

\bibitem[\protect\citeauthoryear{Cordes, Rankin \& Backer}{Cordes
  et~al.}{1978}]{crb78}
Cordes J.~M.,  Rankin J.~M.,    Backer D.~C.,  1978, ApJ, 223, 961

\bibitem[\protect\citeauthoryear{Costa, McCulloch \& Hamilton}{Costa
  et~al.}{1991}]{cmh91}
Costa M.~E.,  McCulloch P.~M.,    Hamilton P.~A.,  1991, MNRAS, 252, 13

\bibitem[\protect\citeauthoryear{{Dyks}, {Rudak} \& {Harding}}{{Dyks}
  et~al.}{2004}]{drh04}
{Dyks} J.,  {Rudak} B.,    {Harding} A.~K.,  2004, ApJ, 607, 939

\bibitem[\protect\citeauthoryear{{Everett} \& {Weisberg}}{{Everett} \&
  {Weisberg}}{2001}]{ew01}
{Everett} J.~E.,  {Weisberg} J.~M.,  2001, ApJ, 553, 341

\bibitem[\protect\citeauthoryear{{Gangadhara} \& {Gupta}}{{Gangadhara} \&
  {Gupta}}{2001}]{gg01}
{Gangadhara} R.~T.,  {Gupta} Y.,  2001, ApJ, 555, 31

\bibitem[\protect\citeauthoryear{Gould \& Lyne}{Gould \& Lyne}{1998}]{gl98}
Gould D.~M.,  Lyne A.~G.,  1998, MNRAS, 301, 235

\bibitem[\protect\citeauthoryear{{Hamilton} \& {Lyne}}{{Hamilton} \&
  {Lyne}}{1987}]{hl87}
{Hamilton} P.~A.,  {Lyne} A.~G.,  1987, MNRAS, 224, 1073

\bibitem[\protect\citeauthoryear{Hamilton, McCulloch, Ables \&
  Komesaroff}{Hamilton et~al.}{1977}]{hmak77}
Hamilton P.~A.,  McCulloch P.~M.,  Ables J.~G.,    Komesaroff M.~M.,  1977,
  MNRAS, 180, 1

\bibitem[\protect\citeauthoryear{Hamilton, McCulloch, Manchester, Ables \&
  Komesaroff}{Hamilton et~al.}{1977}]{hmm+77}
Hamilton P.~A.,  McCulloch P.~M.,  Manchester R.~N.,  Ables J.~G.,
  Komesaroff M.~M.,  1977, Nature, 265, 224

\bibitem[\protect\citeauthoryear{{Han}, {Manchester} \& {Qiao}}{{Han}
  et~al.}{1999}]{hmq99}
{Han} J.~L.,  {Manchester} R.~N.,    {Qiao} G.~J.,  1999, MNRAS, 306, 371

\bibitem[\protect\citeauthoryear{{Hotan}, {van Straten} \&
  {Manchester}}{{Hotan} et~al.}{2004}]{hvm04}
{Hotan} A.~W.,  {van Straten} W.,    {Manchester} R.~N.,  2004, Publications of
  the Astronomical Society of Australia, 21, 302

\bibitem[\protect\citeauthoryear{{Karastergiou} \& {Johnston}}{{Karastergiou}
  \& {Johnston}}{2004}]{kj04}
{Karastergiou} A.,  {Johnston} S.,  2004, MNRAS, 352, 689

\bibitem[\protect\citeauthoryear{{Karastergiou}, {Johnston} \&
  {Kramer}}{{Karastergiou} et~al.}{2003}]{kjk03}
{Karastergiou} A.,  {Johnston} S.,    {Kramer} M.,  2003, A\&A, 404, 325

\bibitem[\protect\citeauthoryear{{Karastergiou}, {Johnston} \&
  {Manchester}}{{Karastergiou} et~al.}{2005}]{kjm05}
{Karastergiou} A.,  {Johnston} S.,    {Manchester} R.~N.,  2005, MNRAS, pp
  285--+

\bibitem[\protect\citeauthoryear{{Karastergiou}, {Johnston}, {Mitra}, {van
  Leeuwen} \& {Edwards}}{{Karastergiou} et~al.}{2003}]{kjm+03}
{Karastergiou} A.,  {Johnston} S.,  {Mitra} D.,  {van Leeuwen} A.~G.~J.,
  {Edwards} R.~T.,  2003, MNRAS, 344, L69

\bibitem[\protect\citeauthoryear{{Karastergiou}, {Kramer}, {Johnston}, {Lyne},
  {Bhat} \& {Gupta}}{{Karastergiou} et~al.}{2002}]{kkj+02}
{Karastergiou} A.,  {Kramer} M.,  {Johnston} S.,  {Lyne} A.~G.,  {Bhat}
  N.~D.~R.,    {Gupta} Y.,  2002, A\&A, 391, 247

\bibitem[\protect\citeauthoryear{Kramer}{Kramer}{1994}]{kra94}
Kramer M.,  1994, A\&AS, 107, 527

\bibitem[\protect\citeauthoryear{Lyne \& Manchester}{Lyne \&
  Manchester}{1988}]{lm88}
Lyne A.~G.,  Manchester R.~N.,  1988, MNRAS, 234, 477

\bibitem[\protect\citeauthoryear{McCulloch, Hamilton, Manchester \&
  Ables}{McCulloch et~al.}{1978}]{mhma78}
McCulloch P.~M.,  Hamilton P.~A.,  Manchester R.~N.,    Ables J.~G.,  1978,
  MNRAS, 183, 645

\bibitem[\protect\citeauthoryear{{Malov} \& {Suleimanova}}{{Malov} \&
  {Suleimanova}}{1998}]{ms98a}
{Malov} I.~F.,  {Suleimanova} S.~A.,  1998, Astronomy Reports, 42, 388

\bibitem[\protect\citeauthoryear{Manchester, Hamilton \& McCulloch}{Manchester
  et~al.}{1980}]{mhm80}
Manchester R.~N.,  Hamilton P.~A.,    McCulloch P.~M.,  1980, MNRAS, 192, 153

\bibitem[\protect\citeauthoryear{Manchester, Taylor \& Huguenin}{Manchester
  et~al.}{1975}]{mth75}
Manchester R.~N.,  Taylor J.~H.,    Huguenin G.~R.,  1975, ApJ, 196, 83

\bibitem[\protect\citeauthoryear{{Melrose}}{{Melrose}}{2000}]{mel00a}
{Melrose} D.~B.,  2000, in Kramer M.,  Wex N.,   Wielebinski R.,  eds, Pulsar
  Astronomy - 2000 and Beyond, {IAU} Colloquium 177 The status of pulsar
  emission theory.
Astronomical Society of the Pacific, San Francisco, p.~721

\bibitem[\protect\citeauthoryear{Michel}{Michel}{1991}]{mic91}
Michel F.~C.,  1991, {Theory of Neutron Star Magnetospheres}.
{University of Chicago Press}, Chicago

\bibitem[\protect\citeauthoryear{{Mitra} \& {Li}}{{Mitra} \& {Li}}{2004}]{ml04}
{Mitra} D.,  {Li} X.~H.,  2004, A\&A, 421, 215

\bibitem[\protect\citeauthoryear{Mitra \& Rankin}{Mitra \&
  Rankin}{2002}]{mr02a}
Mitra D.,  Rankin J.~M.,  2002, ApJ, pp 322--336

\bibitem[\protect\citeauthoryear{Radhakrishnan \& Cooke}{Radhakrishnan \&
  Cooke}{1969}]{rc69a}
Radhakrishnan V.,  Cooke D.~J.,  1969, ApL, 3, 225

\bibitem[\protect\citeauthoryear{{Ramachandran}, {Backer}, {Rankin}, {Weisberg}
  \& {Devine}}{{Ramachandran} et~al.}{2004}]{rbr+04}
{Ramachandran} R.,  {Backer} D.~C.,  {Rankin} J.~M.,  {Weisberg} J.~M.,
  {Devine} K.~E.,  2004, ApJ, 606, 1167

\bibitem[\protect\citeauthoryear{Rankin}{Rankin}{1986}]{ran86}
Rankin J.~M.,  1986, ApJ, 301, 901

\bibitem[\protect\citeauthoryear{Rankin}{Rankin}{1993a}]{ran93}
Rankin J.~M.,  1993a, ApJ, 405, 285

\bibitem[\protect\citeauthoryear{Rankin}{Rankin}{1993b}]{ran93b}
Rankin J.~M.,  1993b, ApJS, 85, 145

\bibitem[\protect\citeauthoryear{Ruderman \& Sutherland}{Ruderman \&
  Sutherland}{1975}]{rs75}
Ruderman M.~A.,  Sutherland P.~G.,  1975, ApJ, 196, 51

\bibitem[\protect\citeauthoryear{Stinebring, Cordes, Rankin, Weisberg \&
  Boriakoff}{Stinebring et~al.}{1984}]{scr+84}
Stinebring D.~R.,  Cordes J.~M.,  Rankin J.~M.,  Weisberg J.~M.,    Boriakoff
  V.,  1984, ApJS, 55, 247

\bibitem[\protect\citeauthoryear{Taylor, Manchester \& Lyne}{Taylor
  et~al.}{1993}]{tml93}
Taylor J.~H.,  Manchester R.~N.,    Lyne A.~G.,  1993, ApJS, 88, 529

\bibitem[\protect\citeauthoryear{van Ommen, D'Alesssandro, Hamilton \&
  McCulloch}{van Ommen et~al.}{1997}]{vdhm97}
van Ommen T.~D.,  D'Alesssandro F.~D.,  Hamilton P.~A.,    McCulloch P.~M.,
  1997, MNRAS, 287, 307

\bibitem[\protect\citeauthoryear{von Hoensbroech, Kijak \& Krawczyk}{von
  Hoensbroech et~al.}{1998}]{hkk98}
von Hoensbroech A.,  Kijak J.,    Krawczyk A.,  1998, A\&A, 334, 571

\bibitem[\protect\citeauthoryear{von Hoensbroech \& Xilouris}{von Hoensbroech
  \& Xilouris}{1997a}]{hx97a}
von Hoensbroech A.,  Xilouris K.~M.,  1997a, A\&A, 324, 981

\bibitem[\protect\citeauthoryear{von Hoensbroech \& Xilouris}{von Hoensbroech
  \& Xilouris}{1997b}]{hx97}
von Hoensbroech A.,  Xilouris K.~M.,  1997b, A\&AS, 126, 121

\end{thebibliography}
\end{document}